\documentclass[]{interact}
\usepackage{lmodern}
\usepackage{url}
\usepackage{epstopdf}

\usepackage[natbibapa,nodoi]{apacite}
\theoremstyle{plain}

\theoremstyle{definition}

\theoremstyle{remark}

\usepackage{tabularx}
\usepackage{longtable}

\usepackage[table]{xcolor}
\usepackage{tabularx}
\usepackage{booktabs}

\definecolor{readEasy}{HTML}{d9f0d3}      
\definecolor{readModerate}{HTML}{fddbc7}  
\definecolor{readHard}{HTML}{f4a582}      

\newcommand{\midsize}{\fontsize{8pt}{10pt}\selectfont}

\begin{document}


\title{Analysis of Terms of Service on Social Media Platforms: Consent Challenges and Assessment Metrics}

\author{
\name{Yong-Bin Kang\textsuperscript{a}\thanks{Corresponding author: Yong-Bin Kang. Email: ykang@swin.edu.au} and Anthony McCosker\textsuperscript{a}}
\affil{\textsuperscript{a}Department of Media and Communication, School of Social Sciences, Media, Film and Education, Swinburne University of Technology, Melbourne, Australia}
}

\maketitle

\begin{abstract}

Social media platforms typically obtain user consent through Terms of Service (ToS) presented at account creation, rather than through dedicated consent forms. This study investigates whether consent-related information is clearly communicated within these ToS documents.
We propose and apply a three-dimensional consent evaluation framework encompassing Textual Accessibility, Semantic Transparency, and Interface Design as declared in ToS documents. Using a combination of computational and qualitative analyses, we assess ToS from 13 major social media platforms.
Our findings reveal important shortcomings across platforms, including high linguistic complexity, widespread use of non-committal language, limited disclosure of data retention and sharing practices, and the absence of explicit interface-level commitments to granular or revocable consent. These results indicate that while consent is formally embedded in ToS, it is often presented in ways that constrain clarity and meaningful choice.
Rather than treating ToS documents as informed consent instruments, this study positions them as consent-bearing documents whose design and content shape the conditions under which users are asked to agree to data practices. The proposed framework offers a systematic method for evaluating consent information within ToS in the absence of explicit consent forms and informs the design of clearer, more ethically robust consent mechanisms for data-intensive digital platforms.
\end{abstract}


\section{Introduction} \label{sec:intro}

Social media platforms remain key channels for communication, commerce, social interaction, news and civic discourse for internet users of all ages \citep{datareportal2025users,paljug2025}. Because of this ubiquity and their underlying extractive business models relying on advertising and other forms of information exchange, these platforms raise important governance and ethical concerns about how they collect, process and manage user data. This constitutes one key layer of the power and potential harms that platforms have over their users' lives, and the growing international efforts to improve `platform governance' \citep{Gorwa:2019}. Central to addressing these concerns is the principle of \textit{informed user consent}—an ethical and legal requirement that people fully understand and voluntarily agree to the ways in which their personal data are used \citep{foster2025social, hanlon2023ethical}. This principle is also contextualized as the principle of `privacy self-management', where platform users are required to consent to the extraction and trade in their personal data to gain access to popular platforms \citep{Solove:2013}. User consent is typically embedded in documents such as Terms of Service (ToS)\footnote{In this paper, \textit{Terms of Service (ToS)} is used as an umbrella term encompassing documents variously titled \textit{Terms of Service}, \textit{User Agreements}, or \textit{Terms of Use}. Social media platforms apply these labels inconsistently, but their functional purpose is substantively the same.}. These documents collectively serve as the primary mechanisms through which social media platforms communicate their data practices and obtain user agreement. However, despite criticism over many years \citep{Solove:2013}, user consent mechanisms on most social media platforms are often dense, opaque, and incomprehensible for diverse users \citep{hanlon2023ethical, hartl2025datadonation}. 

While the platform governance principle of privacy self-management relies on ToS-based consent, ToS documents and mechanisms are known to be inadequate, creating unrealistic demands on users. Their lack of clarity and accessibility leads many users to accept terms without fully engaging with them. For instance, a 2023 Pew Research Center survey found that 56\% of American adults frequently click `agree' without reading the content of privacy policies, and an additional 22\% do so sometimes \citep{pew2023privacy}. This means that nearly 8 in 10 users often bypass these documents without reviewing them. This behavior is also often shaped by platform design choices. A 2023 Mercer University report highlighted that companies often employ tactics that discourage users from reading and understanding these agreements through lengthy, complex legal language, and cognitive overload \citep{mercer2023laws}. As a result, users are often left with only simplified options—such as `accept all' or `agree'—delivered via passive interactions like scrolling or checking a box. While these actions satisfy minimal legal requirements, they fall short of ethical standards for voluntariness, transparency, and autonomy \citep{vijayakumar2025illusion, foster2025social}. 

Despite knowledge of the inequalities in autonomy and control created by these consent arrangements, we argue that this remains an important space for holding platforms to account. Evidence of platforms' consent practices can be used to strengthen platform governance policy and regulation. Furthermore, the disconnect—between the ethical ideal of informed consent and the reality of agreement—has serious implications in contexts such as data sharing with increasingly pervasive AI systems and cross-platform profiling. As data practices become more complex and opaque, the burden placed on ToS as the primary vehicle for communicating consent-related information increases. This shift has consequences not only for individual autonomy but also for the legitimacy of data-intensive business models that rely on large-scale personal data collection \citep{hanlon2023ethical, renuka2025privacy}.
This growing misalignment between user expectations and platform practices underscores the need for a systematic evaluation of how consent-related information is presented within ToS documents.

Recent research has focused on legal compliance checking—assessing whether consent mechanisms align with regulatory frameworks such as the General Data Protection Regulation (GDPR) \citep{eu2016gdpr}—which mandates that consent be freely given, specific, informed, and unambiguous \citep{palm2024consent, santos2020cookie}. However, this provides limited insight into whether consent-related information is clearly presented within ToS documents, where users are first required to agree to access the service.

To address this gap, this paper presents a systematic assessment of the content of ToS documents across 13 major social media platforms. Our study evaluates how document structure, language, and interface-level commitments declared in ToS, addressing these research questions:

\begin{itemize}
    \item \textbf{RQ1:} How readable and linguistically accessible are ToS documents presented to users?
    \item \textbf{RQ2:} To what extent do ToS documents clearly and specifically articulate consent-related information?
    \item \textbf{RQ3:} To what extent do ToS documents provide interface-level consent affordances?
\end{itemize}

To answer these questions, we propose a three-dimensional consent evaluation framework that integrates quantitative linguistic analysis with qualitative assessment of document structure and declared interface design features. Building on prior work that introduced readability as a proxy for usability \citep{hanlon2023ethical}, the framework evaluates consent information across three dimensions:

\begin{enumerate}
    \item \textbf{Textual accessibility:} evaluates the readability and linguistic accessibility of ToS documents.  
    \item \textbf{Semantic transparency:} assesses how clearly consent language communicates the nature, purpose, and risks of data collection and processing, with particular attention to vague or non-committal phrasing that undermines interpretability.  
    \item \textbf{Interface design:} evaluates how consent is formally described within ToS documents, including whether agreements declare passive mechanisms (e.g., review-before-agree), bundled consent, or any explicit commitments to revocability.  
\end{enumerate}

This study does not assess user comprehension or behavioral responses to consent mechanisms. Instead, it evaluates whether consent-related information is clearly and systematically presented within ToS documents. 
This paper makes the following contributions:
\begin{itemize}
    \item \textbf{A multi-dimensional consent evaluation framework:} We propose a multi-dimensional framework for evaluating the quality of digital consent as articulated in platform Terms of Service, based on  \textit{Textual Accessibility}, \textit{Semantic Transparency}, and \textit{Interface Design} dimensions.
    \item \textbf{Cross-platform comparative analysis:} We apply this framework to ToS from 13 major social media platforms, identifying systematic patterns and common shortcomings in how consent information is presented.
\end{itemize}

\section{Background and Related Work} \label{sec:related_work}
Understanding how informed consent is defined, implemented, and evaluated across digital environments is critical to assessing the ethical integrity of social media platforms. This section reviews the theoretical and regulatory foundations of informed consent and examines prior studies that have assessed consent practices in social media contexts. 
The discussion has three parts: (1) an overview of the foundational legal and ethical frameworks governing informed consent; (2) key dimensions used in prior work to assess consent mechanisms on social media platforms; and (3) analysis of the limitations of existing assessment approaches that motivate our proposed consent evaluation framework.

\subsection{Foundations of Informed Consent}

Informed consent is a cornerstone of ethical data collection and digital governance, ensuring that individuals understand and voluntarily agree to how their personal data are used. 
Representatively, the GDPR in the European Union \citep{eu2016gdpr}, the Children’s Online Privacy Protection (COPPA) Act in the United States \citep{ftc2013coppa}, and India’s Digital Personal Data Protection (DPDP) Act 2023 \citep{meity2023dpdp} have established informed consent as a legal precondition for the online collection and processing of personal data. 

These regulations require that consent must be freely given, specific, informed, and unambiguous. For example, the GDPR Article~6(1)(a) states that personal data can only be lawfully processed if the data subject has given informed consent for a clearly defined purpose \citep{eu2016gdpr}. COPPA Act further requires verifiable parental consent for data collection from children under 13, along with transparent disclosure of how the data will be used \citep{ftc2013coppa}. The DPDP Act similarly reinforces these standards, requiring that consent be specific, informed, freely given, and revocable at any time, with additional protections for children and data principals \citep{meity2023dpdp}.

Despite the existence of these legal protections, real-world consent practices often fail to meet these standards. Informed consent—originally grounded in biomedical and research ethics—requires that individuals voluntarily agree to data collection or participation based on a clear understanding of what is involved (Chapter~4, `Respect for Autonomy,' in \citealp{rawbone2015biomedical}). In digital contexts, this principle is embedded in data protection regulations such as the GDPR, COPPA, DPDP, and the more recent Digital Markets Act (DMA) \citep{eu2022dma}. However, researchers have shown that consent obtained through dense, obscure, or coercive mechanisms often fails to meet the ethical standard of meaningful informed consent \citep{hanlon2023ethical, schneble2021social}.

While these ethical and legal standards were originally developed for research and biomedical contexts, social media platforms typically operationalize consent through contractual documents (e.g., ToS), raising questions about whether these instruments can adequately support the requirements articulated in this literature.

\subsection{Key Dimensions in Consent Assessment on Social Media Platforms}

Prior research on digital consent and privacy policies has examined a range of dimensions to evaluate how social media platforms communicate data practices and obtain user agreement. 
These dimensions reflect recurring analytic lenses used across regulatory analyses, usability studies, and interface evaluations. Across these strands, five dimensions are particularly emphasized: \textit{readability}, \textit{clarity of language}, \textit{consent mechanism design}, \textit{granularity of choice}, and \textit{revocability of consent}. Readability and clarity affect whether users can reasonably access and interpret consent information. Consent mechanism design, granularity, and revocability concern how consent is structured and whether meaningful choice and ongoing control are supported. Together, these dimensions capture the widely examined shortcomings in social media consent practices identified in the existing literature. Below, we discuss each of these dimensions in detail.

{\textbf{Readability.}}
A study assessed the accessibility of privacy policies from \textit{Meta} and \textit{X} using a combination of readability and reading fluency metrics \citep{hanlon2023ethical}. The study applied six readability formulas—Flesch Reading Ease, Gunning Fog Index, Flesch–Kincaid Grade Level, SMOG Index, Lensear Write Formula, and Coleman–Liau Index—to determine the education level required to understand the policies. Findings indicate that such policies are largely inaccessible to users without tertiary education, undermining their capacity to provide meaningful consent.
The same study also evaluated reading fluency using average reading speeds (128–238 words per minute) to estimate how long it would take users to read the full text. These assessments showed that many users are unlikely to engage with the full content of privacy policies. Long reading times, combined with dense legal language, increase the likelihood of disengagement and uninformed consent.

{\textbf{Clarity.}}
Ethically sound consent should include brief, understandable disclosures, plain-language summaries, and interaction designs that promote active deliberation \citep{schneble2021social}. Among the platforms assessed, \textit{Pinterest} was the only one that provided a simplified, plain-language version of its terms and conditions, revealing the lack of widespread adoption of clarity-focused consent design.

{\textbf{Consent Mechanism Design.}}
Consent mechanisms on major social media platforms—such as \textit{Facebook}, \textit{TikTok}, and \textit{Instagram}—typically rely on passive interfaces, where users click a checkbox or continue using the service without fully reading the terms \citep{schneble2021social}. These mechanisms may meet legal thresholds but often fail to promote informed, voluntary, and deliberative consent, particularly among children and adolescents with limited comprehension of legalistic language. Design patterns such as scroll-to-agree and bundled opt-ins discourage critical engagement and restrict users from tailoring their consent choices.

{\textbf{Granularity.}}
Consent interfaces frequently group multiple data permissions into a single acceptance action, preventing users from selectively opting out of specific practices (e.g., third-party sharing, Ad targeting) \citep{hanlon2023ethical}. For instance, consent is often requested through blanket `accept all' options, review-before-agree flows, or bundled checkboxes—design patterns that limit user autonomy and obscure the specific implications of their consent. This lack of granularity impedes users from exercising selective control over the types of data processing they consent to.

{\textbf{Revocability.}}
The same study \citep{hanlon2023ethical} highlights that most social media platforms provide limited options for users to revisit or modify their consent after initial agreement. Mechanisms for consent withdrawal are typically buried in account settings or not available at all, undermining the notion of consent as an ongoing and revocable process. This lack of accessible revocation pathways  locks users into long-term data processing agreements, even when their preferences or circumstances change. 



Beyond these dimensions of consent assessment, additional concerns are raised regarding transparency of AI-driven data processing \citep{ijcsis2025privacy} and the adequacy of parental consent mechanisms for minors on social media platforms~\citep{Olivia:2024}. 
It has been emphasized that meaningful evaluation of AI-driven data practices often depends on access to technical documentation, system behavior, or audit-level disclosures \citep{Laine:2024, yerby2022deliberately}, while assessment of parental consent compliance necessitates examination of age-gating mechanisms, verification processes, and enforcement practices beyond ToS documents \citep{ohme2024trace}. 

\subsection{Limitations of Consent Assessment Approaches}

Although informed consent is ethically central to data governance, existing approaches to evaluating consent mechanisms on social media platforms remain fragmented and limited in both scope and analytical depth. They frequently assess isolated components—such as the reading complexity of privacy policies or formal compliance with consent requirements—without systematically examining how linguistic structure, semantic clarity, and interface-level consent signalsing jointly shape how consent-related information is presented to users \citep{hanlon2023ethical, schneble2021social}. 

Specifically, these limitations manifest in four recurring gaps. 
First, existing works assess \textit{readability} primarily in privacy policy documents, using readability formulas and, in some cases, reading-time estimates as proxies for consent quality \citep{hanlon2023ethical}. In our study, we apply these measures to ToS documents, where consent is formally obtained at the point of agreement—constrains users’ capacity to engage with consent-related information.

Second, existing assessments often under-theorize \textit{semantic transparency} in consent language. 
Research has shown that many consent documents deliberately employ vague or non-committal language to preserve corporate flexibility and limit legal accountability, thereby obscuring the scope of data practices \citep{yerby2022deliberately, malik2023vagueness}. 
However, prior works have little explored this concern through explicit metrics that distinguish between lexical ambiguity and substantive disclosure.
This paper addresses this gap by evaluating semantic transparency through two complementary lenses: lexical clarity (the presence of vague or non-committal terms) and specificity (the extent to which data practices, user rights, and platform responsibilities are clearly described).

Third, existing literature recognizes the importance of \textit{interface design} in shaping consent behavior, yet lacks systematic methods for assessing how consent is operationalized through interface-level signals~\citep{schneble2021social}. 
Moreover, interface-related concerns are rarely evaluated alongside textual and semantic features of consent documents. This study addresses this gap by introducing a set of defined interface design metrics that capture how consent is framed and signaled in ToS documents.

Fourth, existing research has been often confined to a \textit{narrow} set of platforms. For example, Hanlon and Jones~\citeyearpar{hanlon2023ethical} assess the accessibility of privacy policies for only two platforms, \textit{Meta} and \textit{X}, using a combination of readability and reading fluency metrics. Our study assesses ToS documents of 13 major social media platforms, enabling a broader comparative analysis of how consent is articulated at scale.


\section{Methods}\label{sec:method}

This section presents a multi-dimensional assessment framework to analyze the content and user-facing presentation of ToS documents. The framework evaluates how consent-related information is presented within these documents, focusing on linguistic accessibility, semantic clarity, and interface-level consent signalling.

\subsection{Data}

We selected 13 social media platforms that represent diversity in content types, user demographics, governance models, and corporate ownership structures. This selection includes platforms operated by multinational technology companies as well as decentralized or privacy-oriented services and platforms with strong youth or community engagement. The full set comprises: \textit{BlueSky, Instagram, LinkedIn, Mastodon, Meta, Pinterest, Reddit, Spotify, TikTok, Tumblr, WhatsApp, X}, and \textit{YouTube}.


Data were collected between May-December 2025 through a manual web-scraping process. For each platform, the following primary user-facing legal documents were extracted from official websites:  
\textit{Terms of Service, Terms of Use}, or \textit{User Agreements}.  
These documents represent the core mechanisms through which platforms communicate data handling practices and obtain consent.
Table~\ref{tab:platform_urls} (Appendix~\ref{appendix:extra_info}) lists their URLs.

\subsection{Multi-dimensional consent assessment framework}

The framework consists of three dimensions: 1) \textit{Textual Accessibility}: assessing readability and reading fluency; 2) \textit{Semantic  Transparency}: evaluating clarity and precision in how consent-related terms are expressed; and 3) \textit{Interface Design}: analyzing how the act of consent is operationalized. We discuss each dimension in detail below.

\paragraph*{Textual Accessibility.}
Although textual accessibility does not directly evaluate consent decisions, it constitutes a foundational condition for document-based consent. Because ToS documents are consent-bearing instruments, users must be able to read and cognitively process their content before consent-related information can be meaningfully understood. The \textit{Textual Accessibility} dimension therefore assesses the extent to which ToS documents are linguistically accessible to users with varying levels of literacy and digital experience.

To operationalize this dimension, we apply seven readability metrics \citep{hanlon2023ethical, siteimprove2024readability}  alongside an estimation of reading fluency based on average reading speeds \citep{brysbaert2019reading} for each ToS document. Together, these eight measures provide a document-level assessment of the cognitive effort required to access and comprehend consent-relevant information.

\begin{table}[ht!]
\small
\centering
\caption{Readability metrics for assessing readability in ToS documents. 
}
\label{tab:readability}
\setlength{\tabcolsep}{5pt} 
\scalebox{0.9}{
\begin{tabularx}{\textwidth}
{
  >{\raggedright\arraybackslash}p{3cm}  
  >{\raggedright\arraybackslash}p{6.5cm}  
  >{\raggedright\arraybackslash}p{4,5cm}   
}
\toprule
\textbf{Metric} & \textbf{Formula} & \textbf{Interpretation (U.S. Education System)} \\
\midrule
Flesch Reading \newline Ease & $206.835 - 1.015 \times (\text{words/sentences}) - 84.6 \times (\text{syllables/words})$; assesses sentence length and syllables per word. & 90–100: Very easy (5th grade); \newline 60–70: Standard (8th–9th); \newline 0–30: Very difficult (college graduate) \\ \addlinespace[10pt]
Gunning Fog \newline Index & $0.4 \times [(\text{words/sentences}) + 100 \times (\text{complex words/words})]$; estimates years of formal education needed to understand text. & $\le$8: Middle school; \newline $\ge$10: High school; \newline $\ge$12: College level \\ \addlinespace[10pt]
Flesch–Kincaid \newline Grade Level & $0.39 \times (\text{words/sentences}) + 11.8 \times (\text{syllables/words}) - 15.59$; converts text complexity into a grade level. & $\le$8: Easy; \newline 9–12: High school; \newline $\ge$13: College \\ \addlinespace[10pt]

Coleman–Liau \newline Index & $0.0588 \times L - 0.296 \times S - 15.8$; based on average letters per 100 words ($L$) and sentences per 100 words ($S$). & $\le$8: Middle school; \newline $\ge$10: High school; \newline $\ge$13: College \\ \addlinespace[10pt]
SMOG Index & $1.0430 \, \sqrt{\text{Polysyllables} \times \frac{30}{\text{Sentences}}} + 3.1291$; predicts grade level using frequency of complex words. & $\le$8: Easy; \newline $\ge$10: Hard; \newline $\ge$13: Academic \\ \addlinespace[10pt]
Lensear Write \newline Formula & $[(\text{easy word count} \times 1) + (\text{hard word count} \times 3)]/{\text{sentence count}}$ &    
$\leq 9$: Middle school;\newline
10--12: High school;\newline
$\geq 13$: College level;\\ \addlinespace[10pt]
Automated \newline  Readability Index (ARI) & $4.71 \times (\text{characters/words}) + 0.5 \times (\text{words/sentences}) - 21.43$; useful for screen-based texts. & $\le$9: Middel School; \newline $\ge$11: High School; \newline $\ge$13: College \\
\bottomrule
\end{tabularx}
}
\end{table}

Table~\ref{tab:readability} summarizes the formulas, scoring ranges, and standard interpretation thresholds. The seven readability formulas estimate the educational level required for comprehension based on variables—sentence length, word complexity, syllables per word, and total characters. 
The {reading fluency} is estimated using average words-per-minute (WPM) benchmarks drawn from educational and linguistic studies \citep{brysbaert2019reading}, which indicates a sense of the time burden. 
For each ToS document, the total word count was divided by these WPM values, categorizing three reader groups, as seen in Table \ref{tab:reading_speed}.  

\begin{table}[ht!]
\small
\centering
\caption{Reading speed benchmarks used for estimating reading fluency.}
\label{tab:reading_speed}
\setlength{\tabcolsep}{8pt} 
\scalebox{0.95}{
\begin{tabular}{>{\raggedright\arraybackslash}l  >{\raggedright\arraybackslash}l}
\toprule
\textbf{Reader Group} & \textbf{Average Reading Speed} \\
\midrule
Children (Age 12–13, Oral) & 120–128 WPM\\
Average Adult (Oral) & $\sim$183 WPM\\
Average Adult (Silent) & $\sim$238 WPM\\
\bottomrule
\end{tabular}
}
\end{table}

\paragraph*{Semantic Transparency.}\label{sec:semantic_transparency}
The \textit{Semantic Transparency} dimension evaluates whether  language in ToS documents is clear and precise enough to enable informed consent. It focuses on how transparently ToS documents describe data practices, user rights, and platform responsibilities. Prior research has shown that many platform privacy policies and ToS frequently rely on vague or non-committal language to preserve corporate flexibility and reduce legal accountability, highlighting that lexical imprecision can obscure the scope of data practices \citep{yerby2022deliberately}. Similarly, another study found that vague terms are not just common—they are systematically embedded to exploit cognitive ambiguity and diminish transparency \citep{malik2023vagueness}.

Specifically, we assess semantic transparency using two metrics, as effective interpretation of platform obligations requires both lexically clear language and sufficiently specific articulation of data practices and user rights:

\begin{itemize}
    \item \textbf{Lexical clarity:} assesses the prevalence and contextual use of vague or non-committal terms (e.g., `may', `from time to time', `as necessary') that weaken the interpretability of platform obligations and user rights.
    \item \textbf{Specificity:} evaluates how precisely ToS documents describe data collection, use, sharing, and retention practices, distinguishing concrete disclosures from abstract or generalized formulations (e.g. `data may be shared with trusted partners'). 
\end{itemize}

To evaluate lexical clarity, we used 30 vague or non-committal terms commonly found in ToS and privacy policy documents. These terms were selected based on our exploratory analysis of prior studies, where such language is frequently used to describe data practices and platform obligations \citep{lebanoff2018, wiggers2023contracts}.
Automated detection is used to identify occurrences of these terms. For each term, we also include similar terms that may be used interchangeably to convey comparable ambiguity or legal flexibility. Full term lists and examples are provided in Table~\ref{tab:vague_terms} (Appendix~\ref{appendix:extra_info}).

The second metric, \textit{Specificity}, evaluates the extent to which ToS documents communicate data practices in concrete terms, drawing on regulatory guidance that requires consent to be informed, specific, and clearly articulated, avoiding undefined or broad descriptions of data handling practices~\citep{oaic_consent_personal_information, ipc_nsw_fact_sheet_consent}.
Low specificity ToS documents---for example, `some information', `affiliates', or stating that data `may be retained as necessary'---may satisfy minimal legal disclosure requirements but provide limited practical understanding for users. In contrast, high-specificity ToS documents explicitly name the data types being collected, identify recipient entities, specify retention periods, and describe concrete purposes for data use and sharing.
To estimate specificity, we use four metrics, summarized in Table~\ref{tab:specificity_metrics}. Each metric is scored on a 0--2 ordinal scale. The scores of these metrics are then combined to form a composite \emph{specificity score}. A full description of the detection rules and implementation details is provided in Appendix~\ref{appendix:specificity}.

\begin{table}[h!]
\midsize
\centering
\caption{The four metrics used to assess the {Specificity} dimension.}
\label{tab:specificity_metrics}
\scalebox{0.9}{
\begin{tabularx}{\textwidth}
{
  >{\raggedright\arraybackslash}p{2cm}
  >{\raggedright\arraybackslash}p{2.5cm}
  >{\raggedright\arraybackslash}p{2.8cm}
  >{\raggedright\arraybackslash}p{2.8cm}
  >{\raggedright\arraybackslash}p{3cm}
}
\toprule
\textbf{Metric} & \textbf{Definition} & \textbf{Evaluation Focus} & \textbf{Assessment Method} & \textbf{Scoring} \\
\midrule
{Named Data Types} & Identifiable categories of data collected. & Are concrete data types (e.g., IP address, location, device ID) explicitly named? & Count distinct concrete data types versus generic labels. &
0 = none \newline
1 = few (1--3 types) \newline
2 = several ($\geq 4$ types)\\
\addlinespace[10pt]

{Named Entities} & Disclosure of specific organizations that process data. & Does the policy name companies instead of using only general terms (e.g., `third parties')? & Detect occurrences of named organizations or corporate suffix patterns. &
0 = none \newline
1 = partial (1--2 entities) \newline
2 = many ($\geq 3$ entities)  \\
\addlinespace[10pt]

{Retention Detail} & Clarity of how long data is kept. & Are numeric or time-bound retention periods specified? & Identify explicit durations. &
0 = none \newline
1 = vague only \newline
2 = explicit duration present \\
\addlinespace[10pt]

{Sharing Conditions} & Justifications for data use or sharing. & Are the reasons or triggers for sharing clearly described? & Detect causal or purpose clauses and classify them as generic or specific. &
0 = none \newline
1 = generic only or 1-2 specific \newline
2 = several specific purposes ($\geq 3$)  \\
\bottomrule
\end{tabularx}
}
\end{table}

\paragraph*{Interface Design.}
The \emph{Interface Design} dimension evaluates how consent-related information is presented  through the structure and interaction flow of the user interface within ToS documents. While legal texts specify the content of consent, interface design determines how consent is enacted, shaping whether it is experienced as informed, voluntary, and unambiguous.  Poorly designed interfaces can undermine user agency, making it easier to accept `agree' than to critically engage with the consent process \citep{utz2019uninformed}. In contrast, a well-designed interface improves users to know what they are consenting to, how they are doing so, and how they can reverse it.

Table \ref{tab:interface} shows the five metrics used to assess this dimension. Together, these metrics aim to capture whether the interface requires active user participation, provides alternatives to acceptance, and signals the possibility of managing or withdrawing consent. Each metric is assessed using defined criteria and scored. This dimension does not measure actual user behavior or the outcomes of downstream interactions.

Platforms with high interface design scores enable users to give their intentional, active, and reversible consent. They treat consent not as a procedural hurdle but as a meaningful decision point. Conversely, platforms that score low on these indicators are more likely to rely on coercive or passive consent models, which undermine ethical standards and reduce user trust. Evaluating interface design in this way provides insights into how consent experiences can be improved through clearer, more user-centered interaction design.

\begin{table}[h!]
\midsize
\centering
\caption{The five metrics used to assess the Interface Design dimension.}
\label{tab:interface}
\scalebox{0.9}{
\begin{tabularx}{\textwidth}
{
  >{\raggedright\arraybackslash}p{2.5cm}
  >{\raggedright\arraybackslash}p{2.5cm}
  >{\raggedright\arraybackslash}p{2.5cm}
  >{\raggedright\arraybackslash}p{2.5cm}
  >{\raggedright\arraybackslash}p{3cm}
}
\toprule
\textbf{Metric} & \textbf{Definition} & \textbf{Evaluation Focus} & \textbf{Assessment Method} & \textbf{Scoring} \\
\midrule
{Unticked Checkbox} & Requires user to actively select an option before proceeding. & Does the interface demand a deliberate action? & Check whether a checkbox (or similar) exists and whether it is unticked by default. & 0 = Absent \newline 1 = Present and unticked \\ \addlinespace[5pt]

{Review-Before-Consent} &
Indicates whether the document states or implies that users must review the terms prior to agreement. &
Does the platform declare a procedural constraint on consent? &
Examine whether the text explicitly requires review before agreement. &
0 = Not stated \newline 1 = Stated 
\\ \addlinespace[5pt]

{Separate Consent Steps} & Presents distinct decisions for different data uses or types. & Does the interface allow granular control? & Count how many distinct screens, toggles, or steps are used. & 0 = Single screen \newline 1 = Some separation \newline 2 = Fully disaggregated \\ \addlinespace[5pt]

{Explicit Denial Option} & Provides an actionable `Decline' or `Manage Settings' option. & Are alternatives to acceptance equally accessible? & Check for buttons or links that allow refusal or customization. & 0 = None \newline 1 = Indirect or hidden \newline 2 = Equal visibility \\ \addlinespace[5pt]

{Reversibility Cue} & Indicates that consent can be withdrawn or modified later. & Does the UI offer post-consent control or revocation options? & Look for direct links or messages at the point of consent about managing future changes. & 0 = No mention \newline 1 = Mention only \newline 2 = Link to settings \\

\bottomrule
\end{tabularx}
}
\end{table}

\subsection{Evaluation Methodology}

We use a hybrid evaluation methodology that combines automated computational techniques with human inspection. This approach reflects the nature of consent mechanisms, which involve both textual features that can be systematically analyzed and interface-level elements that require human judgment to improve interpretation.

As the automated methods, NLP techniques are applied to compute readability measures, detecting linguistic patterns, and estimating reading effort across lengthy legal documents. 
Human inspection is incorporated only where contextual interpretation is necessary. In addition, a hybrid approach—where automated detection is followed by targeted human verification—is used to validate selected cases supported by extracted textual evidence. 
Each of the three dimensions of the proposed framework is assessed as follows:

\begin{itemize}
\item \textbf{Automated (Textual Accessibility):} NLP tools are used to compute readability scores, sentence complexity, lexical density, and estimated reading time (Tables~\ref{tab:readability} and \ref{tab:reading_speed}). 
\item \textbf{Hybrid (Semantic Transparency):} Lexical clarity is assessed using fully automated detection of predefined vague or non-committal terms (Table~\ref{tab:vague_terms} in Appendix~\ref{appendix:extra_info}). Given the scale and dispersion of such terms across long ToS documents, no manual verification is performed for this sub-metric.  
Specificity assessment follows a hybrid approach: NLP is used to detect and extract sentences related to data collection, use, sharing, and retention (Table~\ref{tab:specificity_metrics}), after which we manually examine these matched sentences to determine whether disclosures are concrete or generic and assign ordinal scores.
\item \textbf{Manual (Interface Design):} Interface design is assessed through human inspection, manually examining ToS documents to identify evidence for the metrics (Table~\ref{tab:interface}).
\end{itemize}

\section{Results}
This section presents the results of evaluating the ToS documents of 13 major social media platforms using the proposed framework, with findings reported across the three dimensions outlined in Section~\ref{sec:method}: Textual Accessibility, Semantic Transparency, and Interface Design.  

\subsection{Textual Accessibility}

We report document length and structure, estimated reading-time burden, and readability metrics to provide a multidimensional view of how the textual design of ToS constrains the practical feasibility of informed consent.

\paragraph*{Document Length and Structural Load.}
Before presenting formal readability metrics, we first examine the structural scale of ToS documents, as document length and sentence structure directly shape the cognitive and temporal effort required of users. 

\begin{figure}[h!]
\centering
\includegraphics[width=0.75\textwidth]{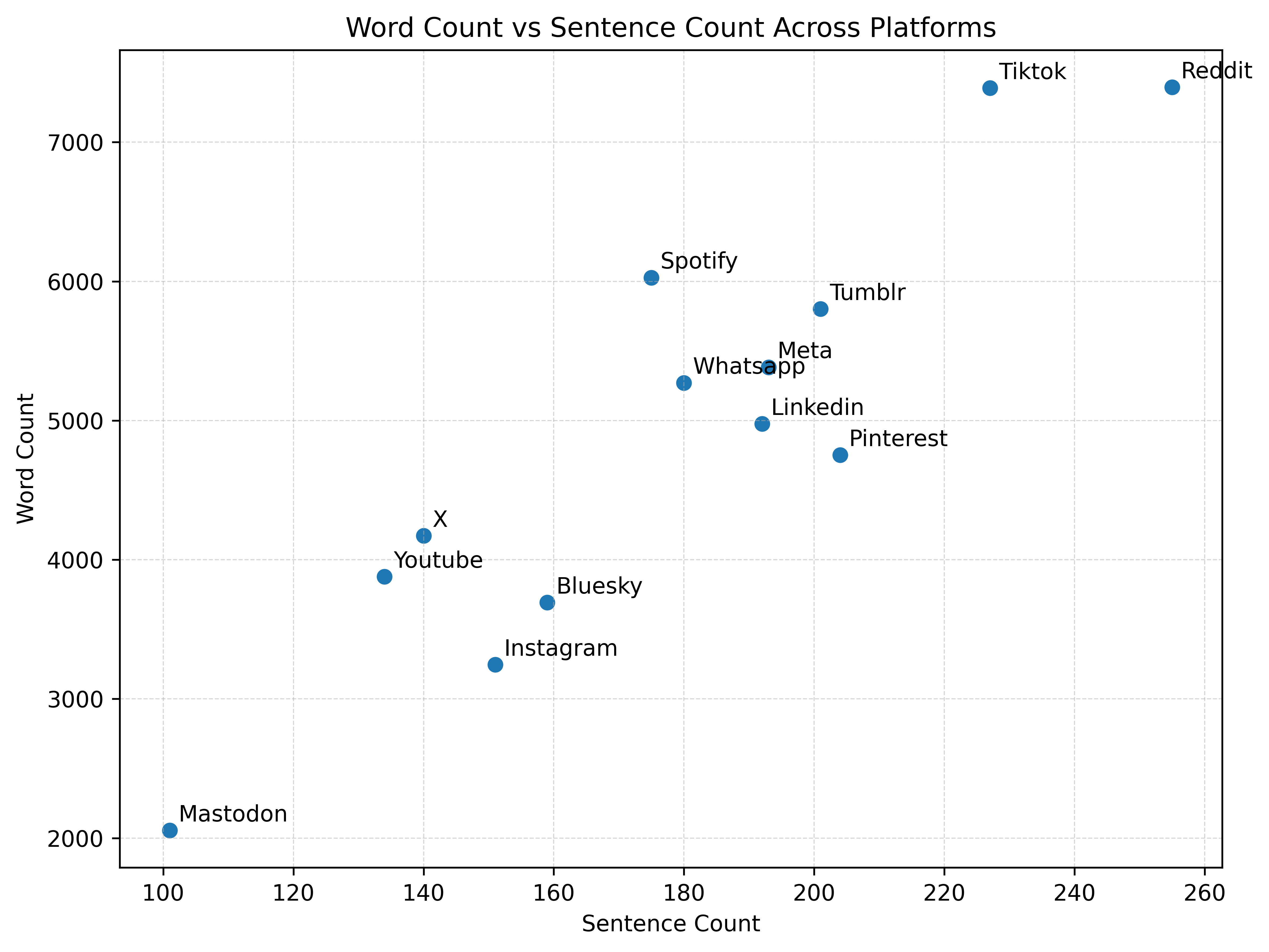}
\caption{Sentence count vs. word count across the 13 ToS documents. 
}
\label{fig:word_sentence_scatter}
\end{figure}

Fig.~\ref{fig:word_sentence_scatter} plots sentence count against word count for each platform’s ToS. Word counts range from 2{,}055 (\textit{Mastodon}) to 7{,}395 (\textit{Reddit}), with sentence counts spanning from 101 to 255. The strong positive association between word and sentence counts indicates that longer ToS documents are not offset by simpler sentence construction. Instead, platforms with greater length also expose users to a higher number of clauses, legal conditions, and nested provisions, compounding both time and cognitive demands.

\begin{figure}[!h]
\centering
\includegraphics[width=0.85\textwidth]{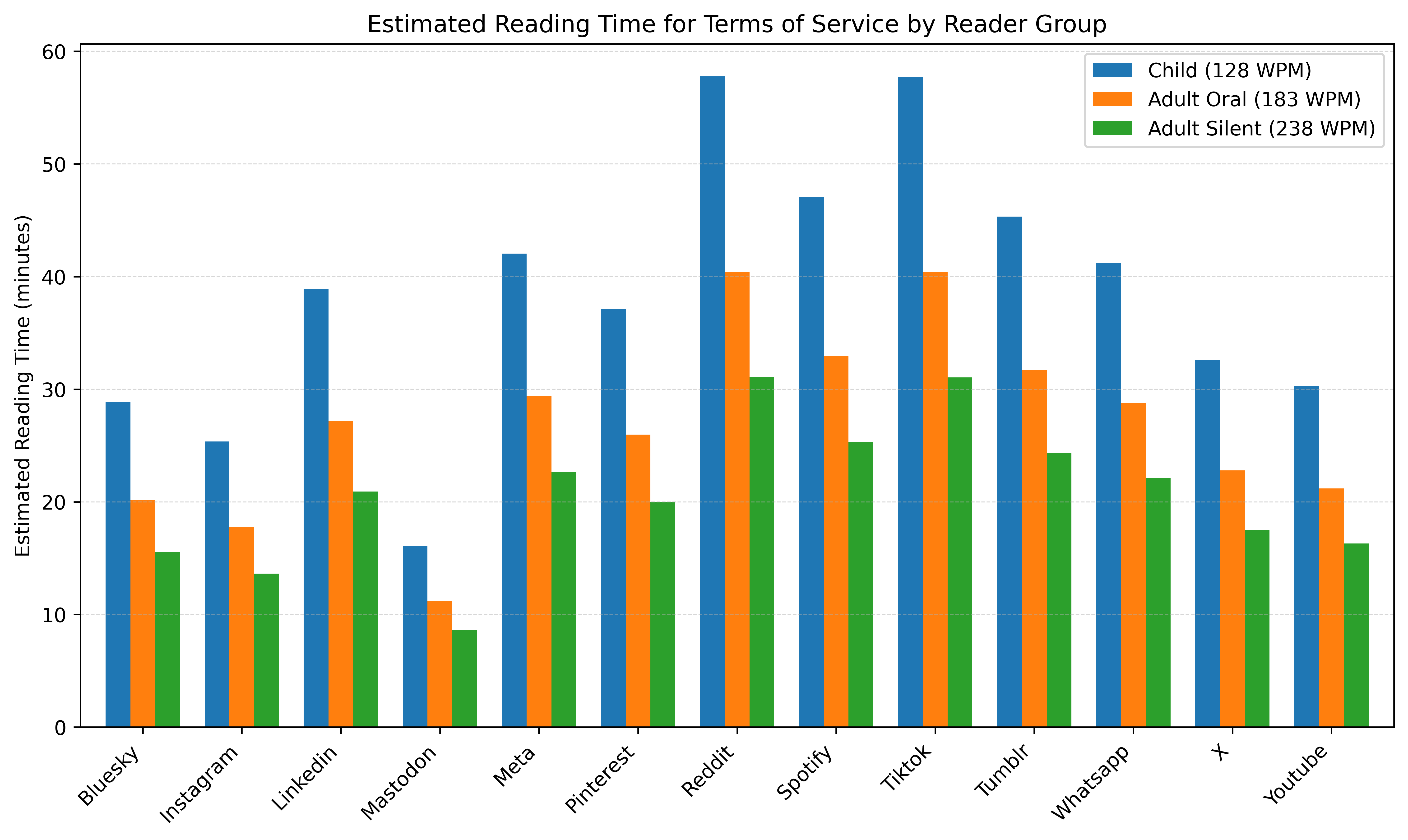}
\caption{Estimated reading time required to read ToS documents.}
\label{fig:reading_time_burden}
\end{figure}

The figure also reveals distinct structural groupings. First,
\textit{Mastodon} forms a clear lower-bound cluster, with both the fewest words (2{,}055) and sentences (101). While comparatively shorter, even this document requires attention and substantial reading (i.e., reading a 2{,}000-word legal document still requires 10-15 minutes reading time for an adult—to be discussed using Fig.~\ref{fig:reading_time_burden}).
Second, an intermediate cluster includes platforms such as \textit{Instagram} (3{,}246/151), \textit{BlueSky} (3{,}696/159), \textit{YouTube} (3{,}879/134), and \textit{X} (4{,}172/140). These platforms maintain moderate document length but still exceed what is typically considered accessible for public-facing informational text.
Third, we observe the group consisting of longer and structurally denser documents, including \textit{LinkedIn} (4{,}976/192), \textit{Meta} (5{,}383/193), \textit{WhatsApp} (5{,}271/180), \textit{Tumblr} (5{,}802/201), and \textit{Pinterest} (4{,}752/204). These platforms combine high word counts with high sentence counts that further elevate reading burden.
Finally, \textit{TikTok} (7{,}398/227) and \textit{Reddit} (7{,}395/255) occupy the extreme upper-right region of the plot, representing the highest structural load in the sample, indicating maximal temporal demands for reading.

Overall, these observations indicate that many platforms require users to engage with documents whose length make full review impractical. 

\paragraph*{Reading Speed Analysis.}

Fig.~\ref{fig:reading_time_burden} shows the time required to read each platform’s ToS.

We observe that even under optimal conditions—adult silent reading at 238 words per minute—most ToS documents require between approximately 15 and 30 minutes to read in full. Platforms such as \textit{Reddit}, \textit{TikTok}, \textit{Tumblr}, and \textit{Spotify} lie at the upper end of this range, with more than 24 minutes. Under average adult oral reading speeds (183 WPM), required reading times largely increase, reaching 20--40 minutes, with \textit{Reddit} and \textit{TikTok} exceeding 40 minutes.
For the children (120--128 WPM), the estimated burden is markedly higher, ranging from 15--20 minutes for the shortest document to more than 55 minutes for the longest one. 

\paragraph*{Readability Analysis.}

Table~\ref{tab:platform_readability} presents the readability profiles of the 13 ToS documents. The results show a pattern of high linguistic complexity, with most documents requiring at least senior secondary or tertiary-level reading proficiency.

\renewcommand{\arraystretch}{1.1} 
\begin{table}[h!]
\small
\centering
\caption{Readability scores: 
Flesch-RE (Flesch Reading Ease), Fog (Gunning Fog Index), F-K Grade (Flesch-Kincaid Grade Level), 
CLI (Coleman-Liau Index), SMOG (SMOG Index), Lensear (Lensear Write Formula), 
ARI (Automated Readability Index). 
Cell colors: 
green: easy or moderate, orange: difficult, red: very difficulty.}
\label{tab:platform_readability}
\begin{tabularx}{\textwidth}{l*{7}{>{\centering\arraybackslash}X}}
\toprule
\textbf{Platform} &
\textbf{Flesch-RE} &
\textbf{Fog} &
\textbf{F-K Grade} &
\textbf{CLI} &
\textbf{SMOG} &
\textbf{Lensear} &
\textbf{ARI} \\
\midrule

BlueSky &
\cellcolor{readModerate}40.2 &
\cellcolor{readHard}16.2 &
\cellcolor{readHard}13.4 &
\cellcolor{readModerate}12.4 &
\cellcolor{readHard}14.9 &
\cellcolor{readModerate}11.3 &
\cellcolor{readHard}14.9 \\

Instagram &
\cellcolor{readModerate}43.4 &
\cellcolor{readHard}15.2 &
\cellcolor{readModerate}12.5 &
\cellcolor{readModerate}11.3 &
\cellcolor{readHard}14.2 &
\cellcolor{readModerate}10.2 &
\cellcolor{readHard}13.2 \\

LinkedIn &
\cellcolor{readModerate}36.4 &
\cellcolor{readHard}17.4 &
\cellcolor{readHard}14.6 &
\cellcolor{readModerate}11.8 &
\cellcolor{readHard}15.6 &
\cellcolor{readHard}13.8 &
\cellcolor{readHard}15.8 \\

Mastodon &
\cellcolor{readModerate}46.5 &
\cellcolor{readHard}13.4 &
\cellcolor{readModerate}11.8 &
\cellcolor{readModerate}10.9 &
\cellcolor{readHard}13.7 &
\cellcolor{readModerate}10.7 &
\cellcolor{readModerate}12.2 \\

Meta (Facebook) &
\cellcolor{readModerate}37.7 &
\cellcolor{readHard}17.3 &
\cellcolor{readHard}14.9 &
\cellcolor{readModerate}11.6 &
\cellcolor{readHard}15.3 &
\cellcolor{readHard}13.8 &
\cellcolor{readHard}16.2 \\

Pinterest &
\cellcolor{readModerate}43.9 &
\cellcolor{readHard}15.1 &
\cellcolor{readModerate}12.9 &
\cellcolor{readModerate}11.1 &
\cellcolor{readHard}13.9 &
\cellcolor{readModerate}11.8 &
\cellcolor{readHard}13.7 \\

Reddit &
\cellcolor{readModerate}31.8 &
\cellcolor{readHard}19.2 &
\cellcolor{readHard}16.0 &
\cellcolor{readModerate}11.9 &
\cellcolor{readHard}16.9 &
\cellcolor{readModerate}10.3 &
\cellcolor{readHard}17.3 \\

Spotify &
\cellcolor{readHard}23.6 &
\cellcolor{readHard}22.1 &
\cellcolor{readHard}18.5 &
\cellcolor{readHard}13.0 &
\cellcolor{readHard}18.7 &
\cellcolor{readHard}24.7 &
\cellcolor{readHard}20.6 \\

TikTok &
\cellcolor{readHard}28.9 &
\cellcolor{readHard}20.3 &
\cellcolor{readHard}17.3 &
\cellcolor{readModerate}12.0 &
\cellcolor{readHard}17.4 &
\cellcolor{readHard}16.8 &
\cellcolor{readHard}18.9 \\

Tumblr &
\cellcolor{readHard}29.7 &
\cellcolor{readHard}19.3 &
\cellcolor{readHard}16.3 &
\cellcolor{readModerate}12.7 &
\cellcolor{readHard}17.0 &
\cellcolor{readEasy}6.9 &
\cellcolor{readHard}17.9 \\

WhatsApp &
\cellcolor{readHard}28.1 &
\cellcolor{readHard}19.6 &
\cellcolor{readHard}16.6 &
\cellcolor{readModerate}12.5 &
\cellcolor{readHard}17.3 &
\cellcolor{readHard}42.5 &
\cellcolor{readHard}17.8 \\

X  &
\cellcolor{readHard}29.0 &
\cellcolor{readHard}19.7 &
\cellcolor{readHard}16.6 &
\cellcolor{readModerate}12.5 &
\cellcolor{readHard}17.3 &
\cellcolor{readHard}22.3 &
\cellcolor{readHard}18.3 \\

YouTube &
\cellcolor{readModerate}36.3 &
\cellcolor{readHard}17.9 &
\cellcolor{readHard}15.4 &
\cellcolor{readModerate}12.0 &
\cellcolor{readHard}15.6 &
\cellcolor{readEasy}5.3 &
\cellcolor{readHard}17.2 \\

\bottomrule
\end{tabularx}
\end{table}

Using grade-based metrics such as Fog, SMOG, and ARI, most platforms fall within the \textit{very difficult} range, typically corresponding to college-level comprehension or higher. Fog scores range from 13.4 (\textit{Mastodon}) to 22.1 (\textit{Spotify}), while SMOG values exceed 13 across all platforms, indicating dense use of polysyllabic vocabulary and complex sentence structures. ARI scores  cluster between 12.2--20.6, reinforcing the conclusion that these texts exceed the reading capacity of the general public.

Some variation appears in metrics that are more sensitive to sentence length and surface-level structure, such as Flesch--RE, F--K Grade, and CLI. For example, \textit{Instagram}, \textit{Mastodon}, and \textit{Pinterest} show scores in the \textit{difficult} band on these measures, indicating that reading level remains at or above late secondary education.

Lensear provides additional pattern. Two platforms—\textit{Tumblr} (6.9) and \textit{YouTube} (5.3)—fall within the \textit{easy} band, suggesting comparatively simpler sentence construction. Nevertheless, this accessibility does not extend across other readability measures for the same platforms, which continue to indicate high overall difficulty. This divergence highlights that isolated improvements in sentence simplicity do not reduce the broader cognitive demands imposed by lengthy, legally dense ToS documents.

Overall, these results emphasize that the ToS documents examined generally fall short of readability levels aligned with recommendations for public-facing informational texts. Even where individual metrics suggest moderate or localized ease, the overall linguistic profile remains dominated by complex syntax and dense informational load. These findings support prior evidence on the opacity of digital consent materials and their textual accessibility barriers~\citep{hanlon2023ethical, boeschoten2022privacy, obar2020}.

To sum up, these findings show that the feasibility of fully reading ToS documents is constrained not only by linguistic complexity but also by time demands. 

\subsection{Semantic Transparency}

Here we present the analysis results for the Semantic Transparency dimension through  two lenses: {lexical clarity} and specificity.

\paragraph*{Lexical Clarity.}
Lexical clarity was assessed by quantifying the occurrence of vague or non-committal terms within each ToS document. Such terms soften platform obligations, obscure responsible actors, or broaden the scope of statements in ways that reduce interpretive certainty for users. Table~\ref{tab:lexical_clarity} shows lexical clarity statistics, including total word count, vague-term count, vague-term density, and the number of unique vague-terms. Detailed results, including the top-15 vague-terms for each platform, are reported in Table~\ref{tab:lexical_clarity_appendix} (Appendix~\ref{appendix:extra_info}).

\begin{table}[!ht]
\scriptsize
\centering
\caption{Lexical clarity statistics, where vague-term density is computed as the proportion of vague or non-committable terms across total document words.}
\label{tab:lexical_clarity}
\begin{tabularx}{0.8\textwidth}{
    l            
    >{\raggedleft\arraybackslash}X  
    >{\raggedleft\arraybackslash}X  
    >{\raggedleft\arraybackslash}X  
    >{\raggedleft\arraybackslash}X  
}
\toprule
\textbf{Platform} & \textbf{Word Count} & \textbf{Vague Term Count} & \textbf{Vague Term Density (\%)} & \textbf{Unique Vague Terms} \\
\midrule
BlueSky & 3{,}779 & 121 & 3.20 & 18 \\
Instagram & 3{,}296 & 138 & 4.19 & 18 \\
LinkedIn & 5{,}073 & 364 & 7.18 & 22 \\
Mastodon & 2{,}088 & 103 & 4.93 & 13 \\
Meta & 5{,}443 & 231 & 4.24 & 20 \\
Pinterest & 4{,}819 & 138 & 2.86 & 19 \\
Reddit & 7{,}503 & 396 & 5.28 & 18 \\
Spotify & 6{,}110 & 218 & 3.57 & 19 \\
TikTok & 7{,}497 & 299 & 3.99 & 17 \\
Tumblr & 5{,}934 & 257 & 4.33 & 21  \\
WhatsApp & 5{,}320 & 271 & 5.09 & 17 \\
X & 4{,}348 & 246 & 5.66 & 21 \\
YouTube & 3{,}923 & 134 & 3.42 & 18 \\
\bottomrule
\end{tabularx}
\end{table}

Vague-term density ranges from 2.86\% (\textit{Pinterest}) to 7.18\% (\textit{LinkedIn}). Even platforms with relatively lower vague-term densities—such as \textit{Pinterest} (2.86\%) and \textit{BlueSky} (3.20\%)—still contain over one hundred ambiguous expressions. At the upper end, \textit{LinkedIn} (7.18\%) and \textit{Reddit} (5.28\%) include several hundred vague or non-committal terms.

Across platforms, vague expressions cluster into three functional categories.: 
(1) \textit{uncertainty language} (e.g., `may,' `might,' `necessary,' `sometimes'), which softens or conditions the commitment of statements; 
(2) \textit{actor ambiguity} (e.g., `third parties,' `third-party,' `others'), which obscures the identities of data recipients; and 
(3) \textit{scope ambiguity} (e.g., `some,' `certain,' `general,' `personal information'), which reduces clarity about the categories of data collected or processed. 
The number of unique vague-terms ranged from 13 (\textit{Mastodon}) to 22 (\textit{LinkedIn}) out of the 30 vague-terms in Table~\ref{tab:vague_terms}, indicating variation not only in frequency but also in the diversity of ambiguous expressions.

\begin{figure}[ht!]
\centering
\includegraphics[width=\textwidth, trim=0cm 3.5cm 0cm 0cm]{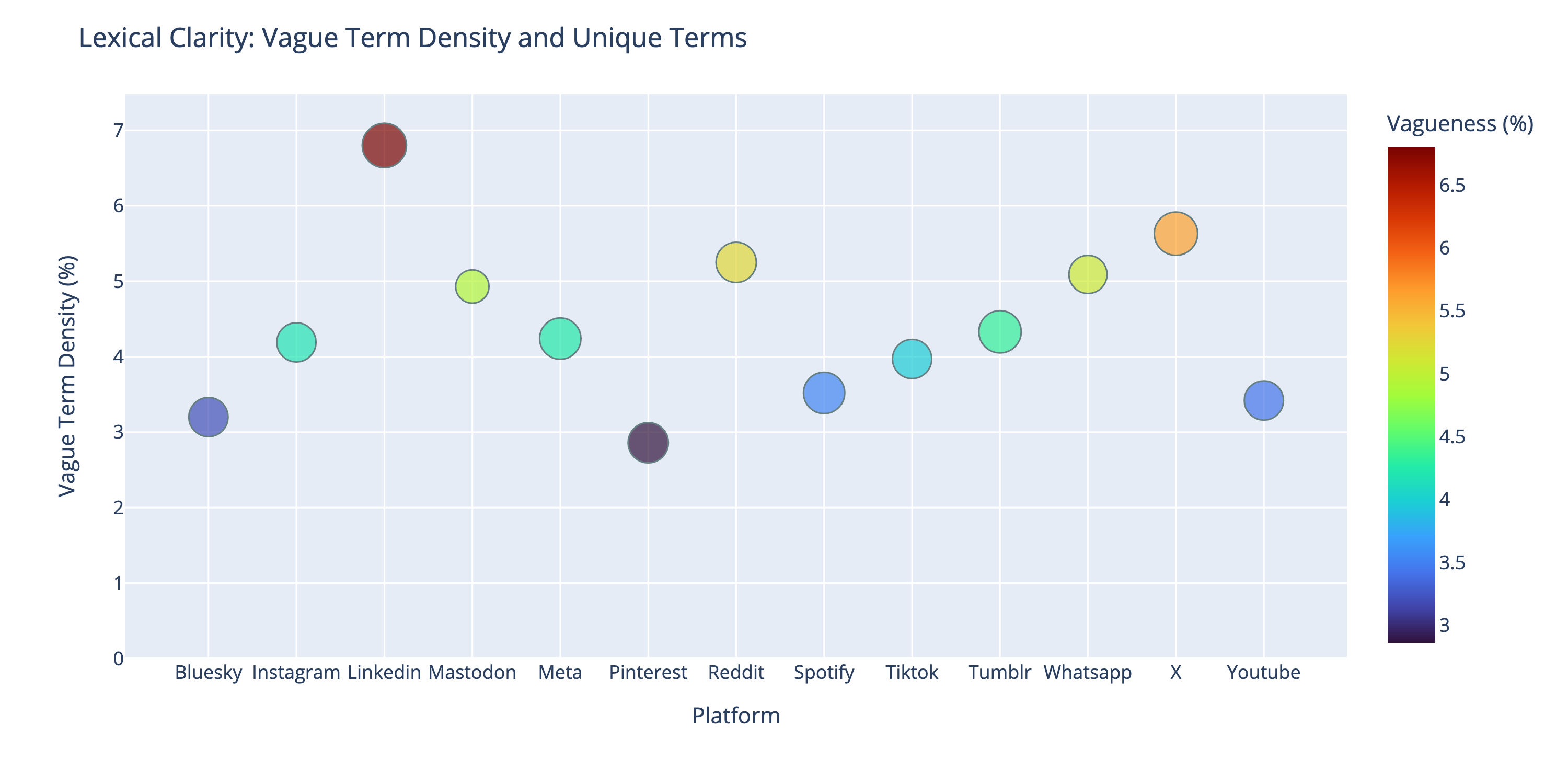}
\caption{Lexical clarity across platforms measured by vague term density and diversity from Table~\ref{tab:lexical_clarity}. }
\label{fig:clarify}
\end{figure}

Fig.~\ref{fig:clarify} shows a comparative view of lexical clarity across platforms by plotting vague-term density (percentage of total words), represented by bubble size (unique vague-terms). These combined dimensions capture how frequently ambiguous language appears, and how varied that language is within each ToS document. Key observations are as follows. First, \textit{LinkedIn} shows the upper-bound case, combining the highest vague-term density with the greatest diversity of vague expressions. Second, a set of \emph{moderate/high-diversity} platforms—\textit{Meta}, \textit{Instagram}, \textit{Reddit}, \textit{TikTok}, \textit{Tumblr},  \textit{WhatsApp}, and \textit{X}—show slightly lower vague-term densities but still draw from a wide range of ambiguous expressions. 
Third, \emph{lower-density, lower-diversity} platforms—such as \textit{Pinterest}, \textit{BlueSky}, and \textit{YouTube}—occupy the lower region. 
Nonetheless, they contain dozens of ambiguous terms. Forth, \textit{Mastodon} shows relatively moderate density and lower diversity.

Collectively, we find that lexical ambiguity is observable across platforms, and their variation indicates differences in both how often vague language is used and how many distinct terms it takes. 

\paragraph*{Specificity.}

Specificity evaluates whether ToS documents describe data practices in concrete and operational terms. While generic expressions such as `personal information', `third parties', or `as long as necessary' may satisfy formal disclosure requirements, they offer limited support for user understanding. 
As stated in Section~\ref{sec:method}, the specificity assessment follows a \textit{hybrid approach}.
Table~\ref{tab:specificity_results} shows both the raw detection counts (e.g., number of distinct data types, entities, retention, and sharing rationales identified) and the resulting ordinal scores: 

\begin{table}[!ht]
\centering
\small
\caption{Document-level specificity metrics. 
Raw counts reflect sentence-level detections (DT = Named Data Types, EN = Named Entities, 
RE = Retention Explicit, SG/SS = Sharing Generic/Specific). 
Scores are shown as \textit{automated $\rightarrow$ post–manual inspection}.}
\label{tab:specificity_results}
\setlength{\tabcolsep}{6pt}
\begin{tabular}{lrrrrrccccr}
\toprule
\textbf{Platform} 
& DT & EN & RE & SG & SS 
& DT$_s$ 
& EN$_s$ 
& R$_s$ 
& S$_s$ 
& \textbf{Specificity} \\
\midrule

Bluesky   
& 10 & 4 & 0 & 0 & 5$\rightarrow$2 
& 2 
& 2 
& 0 
& 1 
& 1.25 \\

Instagram 
& 11 & 4 & 0 & 3$\rightarrow$2 & 5$\rightarrow$3 
& 2 
& 2
& 0
& 2 
& 1.5 \\

LinkedIn  
& 14 & 5 & 0 & 0 & 4$\rightarrow$1
& 2 
& 2 
& 0 
& 1 
& 1.25 \\

Mastodon  
& 6 & 2 & 0 & 0 & 4$\rightarrow$1 
& 2 
& 1
& 0 
& 1
& 1 \\

Meta      
& 15 & 5 & 0 & 5$\rightarrow$3 & 8$\rightarrow$4 
& 2 
& 2
& 0
& 2
& 1.5 \\

Pinterest 
& 13 & 3 & 1$\rightarrow$0 & 1 & 4$\rightarrow$1 
&  2
& 2
& 0
& 1
& 1.25 \\

Reddit    
& 13 & 3 & 0 & 0 & 2$\rightarrow$0 
& 2
& 2
& 0
& 0
& 1 \\

Spotify   
& 8 & 4 & 0 & 0 & 10$\rightarrow$2 
& 2
& 2
& 0
& 1
& 1.25 \\

TikTok    
& 15 & 11 & 0 & 0 & 5$\rightarrow$1 
& 2
& 2
& 0
& 1
& 1.25 \\

Tumblr    
& 13 & 6 & 0 & 0 & 4$\rightarrow$1 
& 2
& 2
& 0
& 1
& 1.25 \\

WhatsApp  
& 13 & 5 & 2 & 6$\rightarrow$1 & 4 
& 2
& 2
& 2
& 2
& 2 \\

X         
& 16 & 1 & 0 & 1 & 2$\rightarrow$0 
& 2 
& 1
& 0
& 1
& 1 \\

YouTube   
& 9 & 5 & 0 & 0 & 3$\rightarrow$0 
& 2
& 2
& 0
& 0
& 1 \\

\bottomrule
\end{tabular}
\end{table}

\textit{Named Data Types:}
All platforms receive a document-level score of~2, detecting multiple named types of data collected. From sentence-level perspectives, platforms include between 6--16 such data types. 
While these sentence-level mentions show acknowledgment of specific named data type, they are often confined to brief or fragmented clauses and are not  accompanied by contextual detail regarding purpose, scope, or retention. For example, \textit{Meta}’s ToS contain multiple types of personal data—such as information about user activity, connections, and interests—but these references are embedded within broad functional descriptions (e.g., advertising or safety) without explicitly mapping individual data types to distinct purposes or retention periods. Similarly, \textit{LinkedIn} enumerates various data types related to personal information in its ToS, yet these mentions are dispersed across licensing, recommendations, and service availability clauses, requiring users to infer how specific data types relate to concrete processing activities or lifecycle constraints. 

\textit{Named Entities:}
Most platforms receive a document-level score of~2, but others rely more heavily on generic descriptors, resulting in scores of~1 (e.g., \textit{Mastodon} and \textit{X}). This variation is observed in sentence-level detections: platforms scoring~2 contain between 3-11 sentences that reference named organizations, whereas \textit{Mastodon} and \textit{X} show fewer such references.
For example, both \textit{Meta} and \textit{LinkedIn} receive sentence-level detections as 5 (EN~=~5), including the mentions of platform names or corporate entities (e.g., Meta companies, LinkedIn affiliates, or legally named corporate parties—`\textit{Meta Platforms, Inc.}', `\textit{Meta Platforms Ireland Limited}', `\textit{LinkedIn Corporation}'), rather than generic phrases like `third parties' or `service providers'. 

\textit{Retention Detail:}
Retention practices show the lowest specificity among the specificity metrics. At the document level, only \textit{WhatsApp} retains a score of~2 after manual inspection, reflecting the presence of multiple explicit and time-bound retention statements. However, other platforms show no sentence-level retention detections after manual validation. \textit{Pinterest} had a single sentence-level retention signal (RE~=~1) during automated analysis; however, manual inspection reveals that this reference relates to timeframes (e.g., dispute or resolution windows). 
Overall, the rarity of retention statements results in uniformly low specificity scores. This pattern highlights a lack of retention-specific transparency, whereby users are rarely informed about how long particular categories of personal data persist, or under what conditions they are deleted.

\textit{Sharing Conditions:}
Sharing conditions show moderate variability across platforms, but manual inspection reveals important differences between procedural clauses and substantive data-sharing disclosures. Several platforms initially received higher sharing scores due to the frequent detection of purpose clauses (e.g., `to provide services' or `to detect/enforce/comply with ...'). 
Manual examination, however, shows that many of these sentence-level detections do not describe the transfer of personal data to identifiable recipients, but instead pertain to contractual enforcement, platform governance, licensing arrangements, or internal operational processes.
For example, \textit{Bluesky} shows 5 sentence-level SS (i.e., Sharing Specific) detections,  which were revised after a manual inspection (SS~=~5$\rightarrow$2; S$_s$=1); \textit{LinkedIn}, \textit{Tumblr}, and \textit{Mastodon} show 4 sentence-level SS detections, which were also reduced to 1 (SS~=~4$\rightarrow$1; S$_s$=1). In these cases, we found that the detected clauses primarily concern platform governance, account enforcement, or content licensing rather than the transfer of personal data to identifiable recipients. \textit{YouTube} and \textit{Reddit} were scored with the result (S$_s$=0): SS~=~3$\rightarrow$0, SS~=~2$\rightarrow$0, respectively.
\textit{WhatsApp} retains a higher-quality sharing profile despite downward revision following manual inspection. Although 6 sentence-level SG detections were initially identified, only 1 survives validation (SG~=~6$\rightarrow$1), with SS~=~4, resulting in S$_s$=2. \textit{Meta} demonstrates relatively strong SS detections: SG~=~5$\rightarrow$3; SS~=~8$\rightarrow$4; S$_s$=2. 
Other platforms, such as \textit{Spotify} and \textit{TikTok}, show moderate sentence-level evidence, resulting in intermediate document-level sharing scores as S$_s$=1.25.
Overall, these results show that while sharing-related language is used in different places at the sentence level, many proportions function at a legal, contractual, or operational level rather than providing users with a clear and concrete account of how their personal data are shared in practice.

In summary, our Specificity assessment shows heterogeneous patterns of specificity across platforms. One platform, \textit{Whatsapp}, achieves the maximum score across all four metrics. However, most ToS combine relatively strong performance on some aspects (most commonly named data types) with limited or absent detail in others, particularly retention practices.
These findings indicate users may receive concrete information on certain aspects of data processing, while remaining insufficiently informed about others, limiting their ability to form a coherent understanding of how their data are handled at the point of agreement.

\subsection{Interface Design}
For the {Interface Design} dimension, our assessment focuses on what platforms formally commit to their contractual texts when users are asked to agree to platform terms. Thus, this assessment captures declared policy commitments (i.e., obligations, permissions, and constraints that platforms articulate in their contractual and policy texts) rather than implemented interface practices (i.e., concrete design and behavior of consent interfaces as they are actually presented to users in situ). 

Our analysis identified the following observations (see also Appendix Table~\ref{tab:interface_analysis_appendix}): 1) None of the ToS documents explicitly describe the use of unticked checkboxes or comparable affirmative interface actions required prior to user agreement. Consent is consistently framed as implicit and conditional on access or continued use of the service; 2) No platform specifies review-before-consent requirements designed to ensure that users review contractual content before agreeing; 
3) Separation of consent decisions is also absent. Although several documents reference multiple policies and describe diverse data uses, user agreement is presented as a single, aggregated action. No platform explicitly commits to presenting users with distinct consent choices for different data uses or purposes. 
4) Denial options are present indirectly. All platforms state that users may refuse the terms by not using the service or by terminating their account. However, refusal is framed as exclusion rather than as a co-equal alternative presented alongside acceptance. As a result, denial exists but as a user-facing choice;
5) Reversibility cues appear frequently, but remain limited in scope. Most ToS documents acknowledge that users may later delete their account or terminate the agreement. These mechanisms are framed as post-consent, high-cost actions and are often accompanied by caveats regarding data retention or delayed deletion. No platform documents interface-level mechanisms enabling users to withdraw or modify consent.
Appendix Table~\ref{tab:interface_analysis_appendix} presents the evidentiary excerpts used to derive all the reesults.

Overall, the Interface Design assessment clearly shows that 
consent is framed as implicit through use, bundled across purposes, reversible through exit, and modifiable—if at all—only after acceptance. 


\section{Discussions} \label{sec:discussion}

This study assessed whether ToS documents of 13 major social media platforms support the conditions required for informed consent. Using a three-dimensional framework—Textual Accessibility, Semantic Transparency, and Interface Design—we evaluated how consent-related information is presented and contractually framed within ToS documents. Our all results converge on a conclusion: although ToS secure formal consent, their structure and operationalization shift an unreasonable cognitive and practical burden onto users, systematically limiting meaningful understanding, deliberation and choice. This goes beyond recognition of the failures of the principle of privacy and data sharing self-management \citep{Solove:2013}. Our results suggest the range of specific mechanisms through which adequate consent is limited by platforms, and hence the key areas in which policy and legislation might target to improve platform governance. The integrated three-dimensional methods for evaluating consent documents provides an opportunity to follow changes over time among platforms, and to evidence policy and governance advocacy efforts.

\textbf{Key Findings:} First, the Textual Accessibility analysis shows that ToS documents are relatively long, linguistically complex, and time-intensive to read, often requiring tertiary-level reading proficiency well beyond realistic user behavior. Second, the Semantic Transparency analysis is undermined by the widespread use of vague and non-committal language and by uneven specificity in describing data practices—particularly with respect to retention and sharing. Third, the Interface Design assessment shows that ToS documents uniformly rely on implicit, bundled, and post-hoc consent structures, with no platform formally committing to active, granular, or low-friction refusal mechanisms.
These findings reveal that while ToS documents function as the primary legal vehicle for consent, they do not sufficiently support the ethical requirements of informed, specific, and voluntary agreement.
These findings also address the three research questions posed in Section~\ref{sec:intro}: (1) addressing \textbf{RQ1}, ToS documents are lengthy, linguistically complex, and time-intensive to read, often exceeding realistic expectations of user engagement; (2) addressing \textbf{RQ2}, consent-relevant information is characterized by pervasive lexical ambiguity and uneven specificity, particularly with respect to data retention and data sharing; and (3) addressing \textbf{RQ3}, ToS documents do not promote active, granular, or low-friction user choice, largely being framed as implicit, bundled, and reversible only through exit.

{\textbf{Implications:}}
Our Textual Readability analysis, addressing \textbf{RQ1}, highlights crucial barriers to user engagement with ToS documents. These documents require advanced reading proficiency and  attention over long periods. The combined effects of document length, high linguistic difficulty, and extended reading-time requirements clearly raise the challenge for most users to review ToS in full before agreeing to them. Under such conditions, acceptance is more plausibly interpreted as a procedural step required for access rather than as an informed and reflective decision.
Our results help explain the widespread tendency to bypass ToS during sign-up and align with empirical evidence showing rapid, non-deliberative consent behavior in digital environments.
From a governance perspective, the findings suggest that the adequacy of consent mechanisms cannot be evaluated on linguistic content alone. Document scale, readability, and time burden can play important role in shaping whether ToS can realistically support informed agreement. Incorporating these factors into consent assessment frameworks is therefore helpful in assessing the ethical and practical validity of platform consent practices.

The Semantic Transparency assessment, addressing \textbf{RQ2},  reveals a dual and consequential pattern: (1) lexical ambiguity is pervasive: even lower-density platforms contain dozens of vague or non-committal expressions. These terms could soften obligations, obscure recipients, and broaden scope, reinforcing findings from prior studies on deliberate opacity in digital contracts; and (2)  specificity is unevenly distributed. Most platforms explicitly name data types, suggesting a baseline level of disclosure. However, this specificity is not matched in other critical areas. Retention practices are rarely described in explicit, time-bound terms, and sharing conditions are frequently overstated by automated detection due to the prevalence of procedural or governance-related clauses. Manual inspection reveals that many apparent `sharing' justifications relate to enforcement, licensing, or service operation rather than substantive data disclosure. Thus, users may gain partial insight into \textit{what} data are collected, while remaining largely uninformed about \textit{how long} data persist or \textit{under what concrete conditions} data are shared. 

The Interface Design assessment, addressing \textbf{RQ2},  focused on examining what platforms formally commit to in their contractual texts. Our findings highlight that no platform explicitly commits to interface mechanisms that require active consent, ensure document review, or present granular choices. 
Our findings align with prior observations that consent interfaces privilege acceptance and continuity over deliberation and refusal, even when legal language nominally preserves user choice.

Collectively, these implications expose that ToS documents are better understood as consent-bearing documents rather than instruments of informed consent. They establish the legal conditions under which consent is presumed, but do not reliably provide the clarity, specificity, or interactional support required for users to understand and sufficiently evaluate data practices at the moment of agreement.

\textbf{Limitations:} This study focuses exclusively on ToS documents referenced and does not assess live interface behavior, post-registration consent settings, or actual user comprehension. Also, the specificity assessment—while strengthened through manual inspection—relies on rule-based detection and ordinal scoring, which cannot capture all nuances of legal interpretation. The analysis is also cross-sectional and does not account for temporal changes in platform policies or regional policy variants. 

\textbf{Future Work:} Future research can extend this framework in several directions. First, integrating interface-level audits and user studies would enable direct comparison between declared policy commitments and actual consent experiences. Second, extending semantic analysis to AI-specific data practices, including training, inference, and cross-context reuse, would address emerging governance challenges. Third, longitudinal analysis of ToS revisions could reveal how platforms adapt consent language in response to regulatory pressure. 
\section{Conclusion}\label{sec:conclusion}

This paper presents a systematic analysis of how consent-related information is articulated within ToS documents of 13 major social media platforms. Specifically, it evaluates whether consent-relevant disclosures are clearly, consistently, and systematically presented within contractual texts that users are asked to accept.
The paper makes two key contributions through the development and application of a unified analytical framework. First, it proposes a multi-dimensional consent evaluation framework that extends beyond readability by integrating textual accessibility, semantic transparency, and declared interface design commitments. Second, it applies this framework in a cross-platform comparative analysis, revealing recurring structural patterns and shared limitations in how consent information is framed, organized and disclosed across contemporary ToS documents.
Our findings indicate that consent-related information in ToS is commonly, largely constrained by a combination of high reading burden, abstract or legally oriented language, uneven specificity—particularly with respect to data retention and sharing—and a lack of explicit contractual commitments to interface-level consent affordances. Collectively, these findings limit the capacity of ToS documents to function as effective vehicles for informed consent. The significance of this paper lies in demonstrating that consent shortcomings are not incidental or platform-specific, but emerge from common design logics embedded in contractual texts that prioritize legal enforceability over user-oriented clarity and deliberation.







\bibliographystyle{apacite}
\bibliography{consent}

\newpage
\appendix
\pagestyle{empty}
\section{Data URLs and Linguistic Clarify Metric}\label{appendix:extra_info}

\begin{table}[h!]
\scriptsize
\centering
\caption{Official URLs used for Terms of Service (ToS) for the 13 platforms analyzed.}
\label{tab:platform_urls}

\begin{tabularx}{\textwidth}{lX}
\toprule
\textbf{Platform} & \textbf{ToS URL} \\
\midrule

BlueSky &
\url{https://bsky.social/about/support/tos}  \\

Instagram &
\url{https://help.instagram.com/581066165581870/}\\

LinkedIn &
\url{https://www.linkedin.com/legal/user-agreement} \\

Mastodon &
\url{https://groups.joinmastodon.org/tos}  \\

Meta &
\url{https://www.facebook.com/legal/terms} \\

Pinterest &
\url{https://policy.pinterest.com/en/terms-of-service} \\

Reddit &
\url{https://www.redditinc.com/policies/user-agreement} \\

Spotify &
\url{https://www.spotify.com/au/legal/end-user-agreement/}\\

TikTok &
\url{https://www.tiktok.com/legal/page/us/terms-of-service/en}  \\

Tumblr &
\url{https://www.tumblr.com/policy/en/terms-of-service} \\

WhatsApp &
\url{https://www.whatsapp.com/legal/terms-of-service}\\

X &
\url{https://x.com/en/tos}  \\

YouTube &
\url{https://www.youtube.com/static?template=terms} \\

\bottomrule
\end{tabularx}
\end{table}

















\renewcommand{\arraystretch}{1.2} 

{\midsize
\begin{longtable}{
  @{}>{\raggedright\arraybackslash}p{0.5cm}
  >{\raggedright\arraybackslash}p{2cm}
  >{\raggedright\arraybackslash}p{5cm}
  >{\raggedright\arraybackslash}p{3cm}
  >{\raggedright\arraybackslash}p{2cm}@{}
}
\caption{Thirty vague or non-committal terms commonly found in privacy policies and user agreements. These terms are frequently used to assess the lexical clarify metric}
\label{tab:vague_terms}\\

\toprule
\textbf{No.} & \textbf{Term} & \textbf{Explanation of Usage} & \textbf{Usage of Terms} & \textbf{Similar Terms} \\ 
\midrule
\endfirsthead

\caption*{\textbf{Table \ref{tab:vague_terms}} (continued)}\\
\toprule
\textbf{No.} & \textbf{Term} & \textbf{Explanation of Usage} & \textbf{Usage of Terms} & \textbf{Similar Terms} \\
\midrule
\endhead

\midrule
\multicolumn{5}{r}{{Continued on next page}}\\
\bottomrule
\endfoot

\bottomrule
\endlastfoot

1 & \textbf{may} & Indicates possibility without obligation. Often used to preserve flexibility in policies and avoid firm commitments. & \textit{We may use the information we collect to personalize features and content.} & \textit{might, could, can} \\ 

2 & \textbf{personal information} & A general term often left undefined or broad. Used to refer to identifiable user data, but what’s included varies across contexts. & \textit{This Privacy Policy explains how we collect, use and share personal information.} & \textit{--} \\

3 & \textbf{information} & Ambiguous blanket term that can refer to any form of user, technical, or behavioural data. Often undefined in scope. & \textit{We collect information about your activity on and …} & \textit{--} \\

4 & \textbf{other} & Used to generalise or imply additional, unspecified items. It obscures what specific data or actors are involved. & \textit{We also receive information about your online and offline actions from other sources.} & \textit{--} \\

5 & \textbf{some} & Indicates an unspecified quantity or subset. Used to soften the perception of extent or coverage. & \textit{Some information is required to use our Products.} & \textit{a few, several, certain} \\

6 & \textbf{certain} & Refers to an undefined subset of data, people, or cases. Offers legal wiggle room and avoids full disclosure. & \textit{Sometimes these integrated partners ask you for permission to access certain additional information from your …} & \textit{selected, specific, unnamed} \\

7 & \textbf{third parties} & Refers to external entities without identifying them. Obscures exactly who will access the user’s data. & \textit{Your data may be shared with third parties for processing.} & \textit{vendors, contractors, partners} \\

8 & \textbf{third party} & Singular form of “third parties.” Used similarly to deflect responsibility or avoid naming recipients. & \textit{We may receive information about you from third party partners.} & \textit{--} \\

9 & \textbf{personally identifiable information} & Technical-sounding phrase often used without listing what qualifies. Creates an illusion of precision while maintaining vagueness. & \textit{We do not sell personally identifiable information without consent.} & \textit{--} \\

10 & \textbf{time to time} & Introduces actions of undefined frequency. & \textit{From time to time, we may update our privacy policy.} & \textit{occasionally, periodically} \\

11 & \textbf{most} & Used to imply majority without exact percentage. Gives impression of normalcy or widespread use while avoiding precise claims. & \textit{Most of the information we collect is used to …} & \textit{--} \\

12 & \textbf{personal data} & Like “personal information,” this is often used broadly and interchangeably, without defining specific categories of data. & \textit{Your personal data is processed in compliance with regulations.} & \textit{--} \\

13 & \textbf{generally} & Indicates a norm with room for exceptions. Common in policy language to hedge statements without full accountability. & \textit{We generally retain data for as long as necessary....} & \textit{usually, typically, in most cases} \\

14 & \textbf{third-party} & Describes entities outside the organization without naming them. Like “third parties,” it is used to abstract and generalize responsibility. & \textit{Some third-party services may have access to your data.} & \textit{external party, vendor} \\

15 & \textbf{others} & Catch-all term to refer to people or entities not listed. Often used to blur or expand the actual audience of data use. & \textit{Data may be disclosed to others ...} & \textit{additional parties, unspecified individuals} \\

16 & \textbf{general} & A vague descriptor that lacks specificity. Used to avoid clear categorization (e.g., “general information” or “general use”). & \textit{IP addresses which we use to estimate your general location…} & \textit{--} \\

17 & \textbf{many} & Indicates a large quantity without being measurable. Allows exaggeration or minimization depending on context. & \textit{Many things influence the content that you see...} & \textit{numerous, several} \\

18 & \textbf{various} & Suggests multiple, often unlisted or changing items. Adds ambiguity to lists or categories. & \textit{We use various tools to collect technical metrics.} & \textit{--} \\

19 & \textbf{might} & Like “may,” it implies possibility without commitment. Often used to sidestep firm statements. & \textit{Your profile might be visible...} & \textit{may, could} \\

20 & \textbf{services} & An all-purpose term often left undefined. May refer to products, features, or business functions, keeping the actual scope unclear. & \textit{We provide services to help people…} & \textit{--} \\

21 & \textbf{non-personal information} & Used to refer to data not directly identifying a person. Often vague due to unclear boundaries between personal and non-personal data. & \textit{We may collect non-personal information.} & \textit{--} \\

22 & \textbf{other information} & A catch-all phrase used to refer to unlisted or miscellaneous data types. Obscures scope by excluding specifics. & \textit{We collect other information such as browser type.} & \textit{--} \\

23 & \textbf{sometimes} & Implies an unspecified, inconsistent action. Used to justify rare or exceptional cases without providing thresholds. & \textit{… sometimes shares the messages that you send to our AIs and general information…} & \textit{occasionally, now and then} \\

24 & \textbf{reasonably} & Used to introduce subjective standards. Implies fairness or effort but lacks objective benchmarks. & \textit{We take reasonably necessary steps to secure your data…} & \textit{appropriately, sensibly} \\

25 & \textbf{appropriate} & Describes actions or decisions without detailing criteria. Used to justify processing or decision-making without transparency. & \textit{We rely on appropriate mechanisms for international data transfers…} & \textit{suitable, fitting, reasonable} \\

26 & \textbf{necessary} & Suggests data processing only happens when justified, but “necessity” is often not clearly defined or externally validated. & \textit{We collect only the data necessary to provide services.} & \textit{required, essential} \\

27 & \textbf{certain information} & Like “certain” and “information,” this compound phrase is doubly vague. It avoids listing the data types involved. & \textit{Certain information may be retained for legal reasons.} & \textit{--} \\

28 & \textbf{typically} & Indicates standard practice but allows for exceptions. Used to reduce commitment while portraying expected behaviour. & \textit{… this will typically include limited information (such as contact details and login information) …} & \textit{generally, customarily} \\

29 & \textbf{affiliates} & Refers to related corporate entities without naming them. Allows broad data sharing without clear disclosure of recipients. & \textit{Your data may be shared with our affiliates…} & \textit{--} \\

30 & \textbf{reasonable} & Used to soften otherwise strict obligations. What counts as “reasonable” is context-dependent and often not legally defined. & \textit{Although we will use reasonable measures to protect your personal data …} & \textit{rational, fair, justifiable} \\

\end{longtable}
}

\begin{table}[!ht]
\scriptsize
\centering
\caption{Lexical clarity statistics for social media Terms of Service, including vague-term density and the top-15 vague expressions per platform.}
\label{tab:lexical_clarity_appendix}
\begin{tabularx}{\textwidth}{
    p{1.1cm}   
    p{1.1cm}   
    p{1.1cm}     
    p{1.2cm}   
    p{1.1cm}   
    >{\raggedright\arraybackslash}X  
}\toprule
\textbf{Platform} & \textbf{Word Count} & \textbf{Vague Term Count} & \textbf{Vague Term Density (\%)} & \textbf{Unique Vague Terms} & \textbf{Top-15 Vague Terms} \\
\midrule
BlueSky & 3{,}779 & 121 & 3.20 & 18 & other, services, may, might, necessary, third party, third-party, general, most, information, some, others, certain, reasonable, many \\
Instagram & 3{,}296 & 138 & 4.19 & 18 & information, may, might, other, others, necessary, services, personal data, certain, reasonable, third parties, many, appropriate, reasonably, affiliates \\
LinkedIn & 5{,}073 & 364 & 7.18 & 22 & services, might, other, information, others, may, necessary, personal data, certain, affiliates, some, third party, third parties, third-party, reasonably \\
Mastodon & 2{,}088 & 103 & 4.93 & 13 & might, other, others, services, information, general, some, reasonably, reasonable, certain, personal data, personal information, necessary \\
Meta & 5{,}443 & 231 & 4.24 & 20 & services, might, other, may, information, others, necessary, personal data, certain, some, third parties, many, reasonable, appropriate, reasonably \\
Pinterest & 4{,}819 & 138 & 2.86 & 19 & other, might, may, necessary, third-party, information, appropriate, third party, services, third parties, others, affiliates, reasonable, various, some \\
Reddit & 7{,}503 & 396 & 5.28 & 18 & services, might, other, information, third-party, necessary, certain, third party, may, affiliates, third parties, some, time to time, appropriate, reasonably \\
Spotify & 6{,}110 & 218 & 3.57 & 19 & might, other, third-party, information, services, third party, certain, reasonable, may, necessary, third parties, time to time, some, others, reasonably \\
TikTok & 7{,}497 & 299 & 3.99 & 17 & services, other, might, third party, information, certain, affiliates, may, third parties, others, reasonable, necessary, general, some, time to time \\
Tumblr & 5{,}934 & 257 & 4.33 & 21 & services, might, other, information, necessary, may, certain, some, third parties, reasonable, affiliates, third party, personal information, others, sometimes \\
WhatsApp & 5{,}320 & 271 & 5.09 & 17 & services, other, might, information, third-party, certain, necessary, others, third parties, may, appropriate, affiliates, some, general, reasonably \\
X & 4{,}348 & 246 & 5.66 & 21 & services, other, might, information, third parties, others, may, certain, necessary, time to time, affiliates, some, third-party, reasonably, most \\
YouTube & 3{,}923 & 134 & 3.42 & 18 & might, other, may, affiliates, some, certain, information, third-party, third party, necessary, services, reasonable, reasonably, others, time to time \\
\bottomrule
\end{tabularx}
\end{table}

\newpage
\section{Specificity Metric Implementation}
\label{appendix:specificity}

This appendix details the implementation of the \textit{Specificity} dimension reported in Section~\ref{sec:semantic_transparency} (see also Tables~\ref{tab:specificity_metrics}). The aim of this metric is to assess how concretely and operationally ToS documents describe data practices that are central to informed consent.

Specificity assessment follows a \textbf{hybrid approach}. Automated natural language processing (NLP) techniques are first used to detect sentences related to data collection, data use, data sharing, and data retention. These automatically matched sentences are then reviewed by manual inspection to determine whether the detected disclosures are substantively informative or merely procedural, legalistic, or governance-oriented. Ordinal scores are assigned at the document level based on this combined assessment.


\subsection*{(1) Lexicons for Named Data Types and Entities}

\paragraph*{Named Data Types.}
To detect explicit references to concrete categories of personal data, we constructed a controlled lexicon of human-interpretable data types. These include, but are not limited to:

\begin{itemize}
    \item \textbf{Identifiers:} \textit{email}, \textit{email address}, \textit{name}, \textit{full name}, \textit{user ID}, \textit{username}, \textit{phone}, \textit{phone number}, \textit{mobile number}.
    \item \textbf{Location and network data:} \textit{IP address}, \textit{IP}, \textit{MAC address}, \textit{location}, \textit{geolocation}, \textit{precise location}, \textit{approximate location}.
    \item \textbf{Device and usage data:} \textit{device ID}, \textit{device identifier}, \textit{device information}, \textit{hardware model}, \textit{operating system}, \textit{OS version}, \textit{browser type}, \textit{browsing history}, \textit{usage data}, \textit{interaction data}, \textit{log data}, \textit{search history}.
    \item \textbf{Payment and billing details:} \textit{payment info}, \textit{payment information}, \textit{billing details}, \textit{billing address}, \textit{payment card}, \textit{credit card}.
    \item \textbf{Media and content:} \textit{profile picture}, \textit{photo/photos}, \textit{video/videos}, \textit{audio}, \textit{voice data}, \textit{posts}, \textit{messages}, \textit{comments}, \textit{reactions}.
    \item \textbf{Contacts and calendar:} \textit{contact list}, \textit{contacts}, \textit{address book}, \textit{calendar events}.
    \item \textbf{Sensitive categories:} \textit{biometric data}, \textit{biometrics}, \textit{health data}.
    \item \textbf{Advertising and profiling identifiers:} \textit{advertising identifier}, \textit{ad ID}, \textit{IDFA}, \textit{GAID}.
\end{itemize}

Sentences containing one or more of these terms are treated as explicitly naming concrete data types.

\paragraph*{Named Entities.}
To capture explicit references to organizations involved in processing, we combined:

\begin{itemize}
    \item a hand-curated list of common platforms and service providers (e.g., \textit{Google}, \textit{Meta}, \textit{Facebook}, \textit{Amazon}, \textit{Apple}, \textit{Microsoft}, \textit{Stripe}, \textit{Acxiom}, \textit{Salesforce}, \textit{TikTok}, \textit{YouTube}, \textit{LinkedIn}, \textit{Reddit}, \textit{Twitter/X}, \textit{Pinterest}, \textit{Instagram}, \textit{WhatsApp}, \textit{Mastodon}, \textit{Bluesky}, \textit{Spotify}, \textit{Tumblr}); and
    \item a pattern for generic corporate names, matching an uppercase token followed by a corporate suffix such as \textit{Inc}, \textit{LLC}, \textit{Ltd}, \textit{GmbH}, \textit{SAS}, \textit{Pty}, \textit{Corp}, \textit{Corporation}, or \textit{PLC}.
\end{itemize}

Any explicit mention of such entities is treated as evidence that the policy identifies specific recipients or processors. 

\subsection*{(2) Pattern Families for Retention and Sharing Conditions}

\paragraph*{Retention detail: explicit vs.~vague.}
Retention patterns were split into two groups: \textit{explicit} (time-bound) and \textit{vague} (open-ended or underspecified).

Explicit retention patterns match sentences where a retention-related verb (e.g., \textit{retain}, \textit{store}, \textit{keep}, \textit{preserve}, \textit{maintain}, \textit{delete}) appears in proximity to a numeric time expression (e.g., ``30 days'', ``12 months'', ``5 years''), or where a retention period or schedule is directly associated with a numeric duration. Vague retention patterns match sentences that refer to retention in non-numeric or conditional terms, such as ``as long as necessary'', ``until no longer needed'', or generic references to a retention policy or duration without specifying a timeframe.

When both explicit and vague elements appear in the same sentence, the explicit element is treated as dominant. Retention scoring is presence-based rather than frequency-based, reflecting the fact that retention policies are typically stated only once or twice per document.

\paragraph*{Sharing conditions: generic vs.~specific purposes.}
Sharing-condition patterns detect causal or purpose clauses associated with data use or disclosure. Sentences are initially identified using sharing- or use-related verbs (e.g., \textit{share}, \textit{use}, \textit{process}, \textit{disclose}, \textit{analyze}). Detected sentences are then classified as:

\begin{itemize}
    \item \textbf{Specific sharing justifications}, such as fraud detection, abuse prevention, legal compliance, content moderation, personalization, security enforcement, or explicitly described data exchanges between users, businesses, or corporate entities.
    \item \textbf{Generic sharing justifications}, such as ``to provide our services'' or ``to improve user experience,'' which offer minimal contextual detail; or
\end{itemize}

Simple negation (e.g., ``do not share'') is detected, and purely negative statements are excluded from contributing to sharing scores.

\subsection*{(3) Sentence-level Extraction and Document Aggregation}

Each ToS document is segmented into sentences. For every sentence, the implementation records:

\begin{itemize}
    \item whether it names one or more concrete data types;
    \item whether it names one or more specific external entities;
    \item whether it contains explicit or vague retention information;
    \item whether it contains a generic or specific sharing justification.
\end{itemize}

For named data types and named entities, document-level measures are computed using \textit{sentence coverage}: the proportion of sentences in the document that contain at least one instance of the relevant feature. This normalization reduces sensitivity to document length and verbosity. For retention detail, document-level scoring is presence-based rather than coverage-based, reflecting the fact that retention policies are typically stated once or twice rather than repeatedly. For sharing conditions, aggregation is restricted to sentences that contain sharing- or use-related verbs (e.g., \textit{share}, \textit{use}, \textit{process}, \textit{disclose}). The proportion of such sentences that include explicit purposes is then computed.

\subsection*{(4) Manual Inspection and Score Revision}

Automated detection intentionally over-identifies candidate sentences in order to maximize recall. All sentences contributing to \textit{Retention Detail} and \textit{Sharing Conditions} scores were therefore manually inspected. Manual inspection focuses on the \textbf{functional role} of each sentence. Clauses were excluded or down-weighted if they primarily described:

\begin{itemize}
    \item contractual enforcement or dispute resolution;
    \item platform governance or account termination rights;
    \item content licensing arrangements unrelated to data disclosure;
    \item procedural statements required for service operation without describing data flows.
\end{itemize}

Conversely, sentences were retained as substantively specific when they described concrete data persistence periods, identifiable data flows, direct sharing between users, sharing with business accounts, or sharing within a named corporate group. Revisions resulting from manual inspection are reflected in Table~\ref{tab:specificity_results} using a \textit{before $\rightarrow$ after} notation for the Retention and Sharing sub-scores, as well as for the composite Specificity score.

\subsection*{(5) Mapping to 0--2 Scores and Composite Index}

Document-level measures are mapped to 0--2 specificity scores, aligned with Table~\ref{tab:specificity_metrics}, using fixed, criterion-referenced thresholds rather than relative rankings across platforms:

\begin{itemize}
    \item \textbf{Named Data Types:}
    \begin{itemize}
        \item 0 = no sentences naming concrete data types;
        \item 1 = limited named data types used (1-3 types);
        \item 2 = multiple named data types used ($\geq 4$).
    \end{itemize}

    \item \textbf{Named Entities:}
    \begin{itemize}
        \item 0 = no named entities detected;
        \item 1 = 1-2 distinct entities named, or minimal sentence coverage;
        \item 2 = mulitple distinct entities named ($\geq$3).
    \end{itemize}

    \item \textbf{Retention Detail:}
    \begin{itemize}
        \item 0 = no retention-related statements detected;
        \item 1 = only vague, non-numeric retention language present;
        \item 2 = at least one explicit, time-bound retention statement present.
    \end{itemize}

    \item \textbf{Sharing Conditions:}
    \begin{itemize}
        \item 0 = no detected sharing;
        \item 1 = generic haring only or 1-2 specific sharing expressions;
        \item 2 = multiple specific sharing expressions ($\geq$3).
    \end{itemize}
\end{itemize}

The four sub-scores are averaged to produce an overall \textit{Specificity Score} for each document. This composite score reflects the extent to which a ToS document communicates its data practices in concrete, interpretable, and actionable terms.


\newpage
\section{Interface Design evidentiary excerpts}\label{appedix:interface_design}

\begin{table}[ht!]
\centering
\small
\caption{Interface Design metrics assessed from the 13 ToS documents.}
\label{tab:interface_analysis_appendix}
\begin{tabular}{lp{2cm}p{2cm}p{2cm}p{2cm}p{2cm}}
\toprule
\textbf{Platform} &
\textbf{Unticked-Checkbox} &
\textbf{Review-Before-Consent} &
\textbf{Separate Consent Steps} &
\textbf{Explicit Denial Option} &
\textbf{Reversibility Cue} \\
\midrule
BlueSky     & 0 & 0 & 0 & 1 & 1 \\
Instagram   & 0 & 0 & 0 & 1 & 1 \\
LinkedIn    & 0 & 0 & 0 & 1 & 1 \\
Mastodon    & 0 & 0 & 0 & 1 & 1 \\
Meta        & 0 & 0 & 0 & 1 & 1 \\
Pinterest   & 0 & 0 & 0 & 1 & 1 \\
Reddit      & 0 & 0 & 0 & 1 & 1 \\
Spotify     & 0 & 0 & 0 & 1 & 1 \\
TikTok      & 0 & 0 & 0 & 1 & 1 \\
Tumblr      & 0 & 0 & 0 & 1 & 1 \\
WhatsApp    & 0 & 0 & 0 & 1 & 1 \\
X           & 0 & 0 & 0 & 1 & 1 \\
YouTube     & 0 & 0 & 0 & 1 & 1 \\
\bottomrule
\end{tabular}
\end{table}

{\midsize
\begin{longtable}{
  @{}>{\raggedright\arraybackslash}p{2.2cm}
  >{\raggedright\arraybackslash}p{2.8cm}
  >{\raggedright\arraybackslash}p{9.2cm}
}
\caption{Complete evidentiary excerpts supporting Interface Design indicators.}
\label{tab:interface_evidence_full} \\
\toprule
\textbf{Platform} & \textbf{Indicator} & \textbf{Evidence from Terms of Service} \\
\midrule
\endfirsthead
\toprule
\textbf{Platform} & \textbf{Indicator} & \textbf{Verbatim Evidence from Terms of Service} \\
\midrule
\endhead
\midrule
\multicolumn{3}{r}{\textit{Continued on next page}} \\
\endfoot
\bottomrule
\endlastfoot

Instagram & Unticked Checkbox &
“If you do not agree to these Terms, then do not access or use Instagram.” \\

Instagram & Review-before-Consent &
No language indicating enforced review, locked buttons, or gating mechanisms tied to document navigation. \\

Instagram & Separate Consent Steps &
“These Terms of Use therefore constitute an agreement between you and Meta Platforms, Inc.” \\

Instagram & Explicit Denial Option &
“If you do not agree to these Terms, then do not access or use Instagram.” \\

Instagram & Reversibility Cue &
“You can learn more about how we use information, and how to control or delete your content, review the Privacy Policy and visit the Instagram Help Center.” “If you do not agree to any updated Terms … you can do so by deleting your account.” \\


BlueSky & Unticked Checkbox &
“If you keep using Bluesky Social, you agree to the updated Terms. If you don’t agree, you must stop using Bluesky Social.” \\

BlueSky & Review-before-Consent &
“Please take a moment to carefully read these Terms.” ``In some cases below, we’ve added some summary text in italics to make these Terms easier to navigate…” \\

BlueSky & Separate Consent Steps &
“Your use of Bluesky Social is subject to these Terms of Service (‘Terms’), as well as the Bluesky Social Community Guidelines and Bluesky Social Privacy Policy…”
“User Content is only used only in connection with: (a) providing Bluesky Social… and (b) promoting and marketing Bluesky and Bluesky Social.” \\

BlueSky & Explicit Denial Option &
“If you don’t agree, you must stop using Bluesky Social.” “To use Bluesky Social, you must create an account (‘Account’).”\\

BlueSky & Reversibility Cue &
“If you want to delete your Account, you can do so through your account settings.” 
“Given the design of the open network, it is not guaranteed that Bluesky will be able to do so comprehensively.” \\


LinkedIn & Unticked Checkbox &
“When you use our Services you agree to all of these terms.” 
“By creating a LinkedIn account or accessing or using our Services … you are agreeing to enter into a legally binding contract with LinkedIn.” \\

LinkedIn & Review-before-Consent &
“If you do not agree to this contract (‘Contract’ or ‘User Agreement’), do not create an account or access or otherwise use any of our Services.” “As a Visitor or Member of our Services, the collection, use, and sharing of your personal data is subject to our Privacy Policy, our Cookie Policy and other documents referenced in our Privacy Policy, and updates.”\\

LinkedIn & Separate Consent Steps &
“Your use of our Services is also subject to our Cookie Policy and our Privacy Policy…” \\

LinkedIn & Explicit Denial Option &
“If you do not agree to this contract (‘Contract’ or ‘User Agreement’), do not create an account or access or otherwise use any of our Services.” “If you wish to terminate this Contract at any time, you can do so by closing your account and no longer accessing or using our Services.”\\

LinkedIn & Reversibility Cue &
“If you wish to terminate this Contract at any time, you can do so by closing your account…” ``“Both you and LinkedIn may terminate this Contract at any time with notice to the other. On termination, you lose the right to access or use the Services.” \\


Mastodon & Unticked Checkbox &
“Everyone needs to agree to these terms to use the forum.” “Once you get notice of an update to these terms, you must agree to the new terms in order to keep using the forum.” \\

Mastodon & Review-before-Consent &
“When you use our Services you agree to all of these terms.” “If you do not agree to this contract (‘Contract’ or ‘User Agreement’), do not create an account or access or otherwise use any of our Services.”\\

Mastodon & Separate Consent Steps &
 ``Subject to these terms, the company gives you permission to use the forum.” “Your permission to use the forum is subject to the following conditions:”\\

Mastodon & Explicit Denial Option &
 “Everyone needs to agree to these terms to use the forum.” “Either you or the company may end the agreement written out in these terms at any time.”\\

Mastodon & Reversibility Cue &
 “You may close your account at any time by e-mailing $<$contact\_email$>$.” “Either you or the company may end the agreement written out in these terms at any time. When our agreement ends, your permission to use the forum also ends.” \\


Meta & Unticked Checkbox &
“If you do not agree to these Terms, then do not access or use Facebook or the other products and services covered by these Terms.” “By using our Products, you agree that we can show you ads that we think may be relevant to you and your interests.”\\

Meta & Review-before-Consent &
“These Terms govern your access and use of Facebook, Messenger and the other products, websites, features, apps, services, technologies and software we offer.” “If you do not agree to these Terms, then do not access or use Facebook or the other products and services covered by these Terms.”\\

Meta & Separate Consent Steps &
 ``By using our Products, you agree that we can show you personalized ads and other commercial and sponsored content.”  “We use your personal data to help determine which personalized ads to show you.”\\

Meta & Explicit Denial Option &
 “If you do not agree to these Terms, then do not access or use Facebook or the other products and services covered by these Terms.” “If you do not agree to our updated Terms, or wish to terminate your agreement to this contract, you can delete your account at any time and you must also stop accessing or using Facebook and the other Meta Products.”\\

Meta & Reversibility Cue &
 “You can delete individual content that you share, post and upload at any time.”
“If you delete your account, and you stop accessing, using or visiting Facebook and the other Meta Products, or if this contract is otherwise terminated, then these terms shall terminate as an agreement between you and us.” “It may take up to 90 days to delete content after we've begun the account deletion process.” \\


Pinterest & Unticked Checkbox &
“By accessing or using Pinterest, you agree to comply with and be bound by these Terms…” “If you do not agree to our Terms, you must not access or use Pinterest.”\\

Pinterest & Review-before-Consent &
“Please read these Terms carefully and contact us if you have any questions.” “More simply put”\\

Pinterest & Separate Consent Steps &
 “For clarity, these Terms include, and incorporate by reference, the following policies…”
“By accessing or using Pinterest, you agree to comply with and be bound by these Terms…”\\

Pinterest & Explicit Denial Option &
“If you do not agree to our Terms, you must not access or use Pinterest.” “You can also terminate or delete your account at any time.”\\

Pinterest & Reversibility Cue &
 “You can also terminate or delete your account at any time.”
“Following termination or deactivation of your account, or User Content removal from Pinterest, we may keep your User Content for a reasonable period of time…”\\


Reddit & Unticked Checkbox &
“By accessing or using our Services, you agree to be bound by these Terms.”  “If you do not agree to these Terms, you may not access or use our Services.”\\

Reddit & Review-before-Consent &
“By accessing or using our Services, you agree to be bound by these Terms.” “Remember Reddit is for fun and is intended to be a place for your entertainment, but we still need some basic rules.”\\

Reddit & Separate Consent Steps &
“These Terms, together with the Privacy Policy and any other agreements expressly incorporated by reference into these Terms, constitute the entire agreement between you and us regarding your access to and use of the Services.” “You understand that through your use of the Services, you consent to the collection and use of this information as set forth in the Privacy Policy.”\\

Reddit & Explicit Denial Option &
“If you do not agree to these Terms, you may not access or use our Services.” “You may terminate these Terms at any time and for any reason by deleting your Account and discontinuing use of all Services.”
\\

Reddit & Reversibility Cue &
 “You may terminate these Terms at any time and for any reason by deleting your Account and discontinuing use of all Services.” “The following sections will survive any termination of these Terms or of your Account…”\\

Spotify & Unticked Checkbox &
“By signing up for, or otherwise using, the Spotify Service, you agree to these Terms.” “If you do not agree to these Terms, then you must not use the Spotify Service or access any Content.”\\

Spotify & Review-before-Consent &
“Please read these Terms of Use (these ‘Terms’) carefully as they govern your use of (which includes access to) Spotify’s personalized services…” “By signing up for, or otherwise using, the Spotify Service, you agree to these Terms.”\\

Spotify & Separate Consent Steps &
“Use of the Spotify Service is subject to additional terms and conditions presented by Spotify, which are hereby incorporated by this reference into these Terms…” “To learn more about how Spotify collects, uses, shares and protects your personal data, please see the Spotify Privacy Policy.”\\

Spotify & Explicit Denial Option &
“If you do not agree to these Terms, then you must not use the Spotify Service or access any Content.” “You may terminate these Terms at any time, in which case you may not continue accessing or using the Spotify Service.”
\\

Spotify & Reversibility Cue &
 “You may terminate these Terms at any time, in which case you may not continue accessing or using the Spotify Service.” “If you disagree with the changes, you can terminate your account before the changes take effect.”\\

Tiktok & Unticked Checkbox &
“By accessing or using our Services, you confirm that you can form a binding contract with TikTok, that you accept these Terms and that you agree to comply with them.” “You can accept the Terms by accessing or using our Services.”\\

Tiktok & Review-before-Consent &
“Please take the time to read them carefully.” “You should print off or save a local copy of the Terms for your records.”\\

Tiktok & Separate Consent Steps &
“Your access to and use of our Services is also subject to our Privacy Policy and Community Guidelines, the terms of which can be found directly on the Platform… and are incorporated herein by reference.”  “By using the Services, you consent to the terms of the Privacy Policy.”\\

Tiktok & Explicit Denial Option &
 “If you do not agree to these Terms, you must not access or use our Services.” “If you no longer want to use our Services again, and would like your account deleted, contact us…”
\\

Tiktok & Reversibility Cue &
 “Once you choose to delete your account, you will not be able to reactivate your account or retrieve any of the content or information you have added.” “Your continued access or use of the Services after the date of the new Terms constitutes your acceptance of the new Terms.”\\

Tumblr & Unticked Checkbox &
“By using or accessing the Services, you agree to be bound by all the terms and conditions of this Agreement.” “If you don't agree to all the terms and conditions of this Agreement, you shouldn't, and aren't permitted to, use the Services.”\\

Tumblr & Review-before-Consent &
 “Please read these Terms of Service and our User Guidelines … carefully before using tumblr.com…” “You should read these Terms carefully before using our site, services, or products.”\\

Tumblr & Separate Consent Steps &
“By using or accessing the Services, you agree to be bound by all the terms and conditions of this Agreement.” “By using the Services you agree you have read the Privacy Policy, which describes our collection, use, and sharing … of such information.”\\

Tumblr & Explicit Denial Option &
 “If you don't agree to all the terms and conditions of this Agreement, you shouldn't, and aren't permitted to, use the Services.” “Either party may terminate this Agreement at any time by notifying the other party.”
\\

Tumblr & Reversibility Cue &
  “You can delete your account at any time here.” “Deleted User Content may persist in caches or backups for a reasonable period of time and … copies of or references to the User Content may not be entirely removed.”\\

Whatsapp & Unticked Checkbox &
“In order to provide our Services … we need to obtain your agreement to our Terms of Service (‘Terms’).” “If you do not agree to our Terms, you must stop using our Services by deleting your account.”\\

Whatsapp & Review-before-Consent &
 “Please read the ‘Special Arbitration Provision For United States Or Canada Users’ section below to learn more.” “Please read these Terms carefully.”\\

Whatsapp & Separate Consent Steps &
“Your use of our Services is subject to these Terms of Service and Privacy Policy.” “By using our Services, you consent to the terms of our Privacy Policy.”\\

Whatsapp & Explicit Denial Option &
 “If you do not agree to our Terms, you must stop using our Services by deleting your account.” “You may stop using our Services at any time.”
\\

Whatsapp & Reversibility Cue &
  “Once you choose to delete your account, you will not be able to reactivate your account or retrieve any of the content or information you have added.” “We may retain certain information even after you have deleted your account.”\\

X & Unticked Checkbox &
“By using the Services you agree to be bound by these Terms.” “You may use the Services only if you agree to form a binding contract with us…”\\

X & Review-before-Consent &
 “You should read these Terms of Service (‘Terms’) in full…” “By using the Services you agree to be bound by these Terms.”\\

X & Separate Consent Steps &
“Please also note that these Terms incorporate our Privacy Policy … as well as other terms applicable to your use of the Services and your Content.” “You understand that through your use of the Services you consent to the collection and use (as set forth in the Privacy Policy) of this information…”\\

X & Explicit Denial Option &
“If you do not agree to these Terms, you should not use the Services.” “You may end your legal agreement with us at any time by deactivating your accounts and discontinuing your use of the Services.”
\\

X & Reversibility Cue &
  “You may end your legal agreement with us at any time by deactivating your accounts and discontinuing your use of the Services.” “By continuing to access or use the Services after those revisions become effective, you agree to be bound by the revised Terms.”\\

Youtube & Unticked Checkbox &
“Please read this Agreement carefully and make sure you understand it. If you do not understand the Agreement, or do not accept any part of it, then you may not use the Service.” “Your use of the Service is subject to these terms…”\\

Youtube & Review-before-Consent &
 “This index is designed to help you understand some of the key updates we’ve made to our Terms of Service (Terms). We hope this serves as a useful guide, but please ensure you read the Terms in full.” “Please read this Agreement carefully and make sure you understand it.”\\

Youtube & Separate Consent Steps &
“Your use of the Service is subject to these terms, the YouTube Community Guidelines and the Policy, Safety and Copyright Policies which may be updated from time to time (together, this ‘Agreement’).” “By using the Service, you agree to this Agreement.”\\

Youtube & Explicit Denial Option &
“If you do not understand the Agreement, or do not accept any part of it, then you may not use the Service.” “You may stop using the Service at any time. Follow these instructions to delete the Service from your Google Account…”
\\

Youtube & Reversibility Cue &
  “You may stop using the Service at any time. Follow these instructions to delete the Service from your Google Account…” “The licenses granted by you continue for a commercially reasonable period of time after you remove or delete your Content from the Service.”\\

\end{longtable}
}

\end{document}